\documentclass[12pt,preprint]{aastex} 

\usepackage{emulateapj5}
\usepackage{epsfig} 
\usepackage{apjfonts}

\newcommand{\nd}{\nodata}
\newcommand{\mdot}{M$_{\odot}$}
\newcommand{\mv}{$M_V$}

\newcommand{\wopt}{$\mathrm{W29_{opt}}$}
\newcommand{\opt}{$\mathrm{_{opt}}$}
\newcommand{\hst}{{\it HST}}
\newcommand{\cha}{{\it Chandra}}

\newcommand{\lx}{$L_{x}$}

\newcommand{\fxfopt}{$F_X/F_{opt}$}
\newcommand{\ergs}{erg~s$^{-1}$}

\shorttitle{Optical Counterparts to 47~Tuc X-ray sources}
\shortauthors{Edmonds et al.}

\begin{document}

\title{An Extensive Census of \hst\ Counterparts to \cha\ X-ray Sources in
the Globular Cluster 47~Tucanae. I. Astrometry and Photometry\altaffilmark{1}}

\altaffiltext{1}{Based on observations with the NASA/ESA {\it Hubble Space
Telescope} obtained at STScI, which is operated by AURA, Inc. under NASA
contract NAS 5-26555.}

\author{Peter D. Edmonds\altaffilmark{2}, Ronald
L. Gilliland\altaffilmark{3}, Craig O. Heinke\altaffilmark{2}, \& Jonathan
E. Grindlay\altaffilmark{2}} \altaffiltext{2}{Harvard-Smithsonian Center
for Astrophysics, 60 Garden St, Cambridge, MA 02138;
pedmonds@cfa.harvard.edu; cheinke@cfa.harvard.edu, josh@cfa.harvard.edu}
\altaffiltext{3}{Space Telescope Science Institute, 3700 San Martin Drive,
Baltimore, MD 21218; gillil@stsci.edu}

\begin{abstract}

We report, in this study of 47~Tucanae, the largest number of optical
identifications of X-ray sources yet obtained in a single globular
cluster. Using deep \cha/ACIS-I imaging and extensive \hst\ studies with
WFPC2 (including a 120 orbit program giving superb $V$ and $I$ images), we
have detected optical counterparts to at least 22 cataclysmic variables
(CVs) and 29 chromospherically active binaries (BY Draconis and RS Canum
Venaticorum systems) in 47~Tuc. These identifications are all based on
tight astrometric matches between X-ray sources and objects with unusual
(non main sequence) optical colors and/or optical variability.  Several
other CVs and active binaries have likely been found, but these have
marginal significance because of larger offsets between the X-ray and
optical positions, or colors and variability that are not statistically
convincing. These less secure optical identifications are not subsequently
discussed in detail.

In the $U$ vs $U-V$ color magnitude diagram (CMD), where the $U$-band
corresponds to either F336W or F300W, the CVs all show evidence for blue
colors compared to the main sequence, but most of them fall close to the
main sequence in the $V$ vs $V-I$ CMD, showing that the secondary stars
dominate the optical light.  The X-ray detected active binaries have
magnitude offsets above the main sequence (in both the $U$ vs $U-V$ or $V$
vs $V-I$ CMDs) that are indistinguishable from those of the much larger
sample of optical variables (eclipsing and contact binaries and BY Dra
variables) detected in the Wide Field Planetary Camera 2 (WFPC2) studies of
Albrow et al. (2001).  We also present the results of a new, deeper search
for optical companions to MSPs. One possible optical companion to an MSP
(47~Tuc~T) was found, adding to the two optical companions already known.
Finally, we study several blue stars with periodic variability from Albrow
et al. (2001) that show little or no evidence for X-ray emission. The
optical colors of these objects differ from those of 47~Tuc (and field)
CVs.

An accompanying paper will present time series results for these optical
identifications, and will discuss X-ray to optical flux ratios, spatial
distributions and an overall interpretation of the results.

\end{abstract}


\section{Introduction}
\label{sect.intro}

The study of globular cluster binaries is motivated by their profound
impact on the dynamical evolution of globular clusters (Hut et
al. 1992). Stellar densities near the centers of globular clusters can
reach values as high as 10$^6$ stars per cubic parsec, and in these extreme
conditions interactions and collisions between single and binary stars
become inevitable (Hurley \& Shara 2002). These interactions can act as a
heating source by the conversion of binary binding energy into stellar
kinetic energy when, for example, low-mass stars are ejected after binary
collisions, helping to stall or prevent core collapse.  Realistic models of
globular cluster evolution therefore require knowledge of their binary star
populations.  

One of the most powerful ways to search for binaries in globular clusters
is by studying their X-ray source population.  The bright (\lx\ $\gtrsim
10^{36}$ erg~s$^{-1}$) X-ray sources in globular clusters have already been
identified with low-mass X-ray binaries (LMXBs; Grindlay et al. 1984) and
the fainter sources (\lx\ $\lesssim 10^{32-33}$ erg~s$^{-1}$) have long
been thought to contain a mixture of LMXBs in quiescence (qLMXBs) and
cataclysmic variables (CVs; Hertz \& Grindlay 1983, Verbunt \& Hasinger
1998). However, secure optical identifications were rare because of the
severe crowding found near the centers of clusters. This problem is
exacerbated by mass segregation, which causes relatively heavy objects like
binaries to sink towards the centers of clusters, where the crowding is at
its worst. Because of these problems, only the brightest binaries in rich
globular clusters like 47~Tucanae were discovered with X-ray missions like
{\it Einstein} and ROSAT.

These resolution limitations have been dramatically overcome by the use of
the \cha\ X-ray observatory. Rich samples of X-ray sources have been
reported in the globular clusters 47~Tuc (Grindlay et al. 2001a; hereafter
GHE01a), NGC~6397 (Grindlay et al. 2001b; hereafter GHE01b), NGC~6752
(Pooley et al. 2002a), NGC~6440 (Pooley et al. 2002b) and $\omega$ Centauri
(Rutledge et al. 2002; Cool, Haggard, \& Carlin 2002). The sample in the
massive globular cluster 47~Tuc is by far the largest, with over 100
sources detected in a 2\arcmin $\times$ 2.5\arcmin\ field centered on the
cluster (GHE01a), and close to 200 sources detected out to a radius of
4\arcmin\ (Grindlay et al. 2002).  These numbers for 47~Tuc alone outnumber
the total sample of known X-ray sources in globular clusters before the
availability of \cha.

With the use of Hubble Space Telescope (\hst) imaging, optical
identifications of X-ray sources in several clusters have been reported
(using both ROSAT and \cha\ X-ray positions), including populations of CVs
in NGC~6397 (Cool et al. 1998 and GHE01b) and NGC~6752 (Bailyn et al. 1996
and Pooley et al. 2002a). In 47~Tuc, Verbunt \& Hasinger (1998) identified
the ROSAT source X9 with the CV candidate V1 (Paresce, de Marchi, \& Ferraro
1992) and X19 with the dwarf nova V2 (Paresce \& de Marchi 1994). Other CV
candidates in 47~Tuc include V3 (Shara et al. 1996) and AKO~9 (Auriere,
Koch-Miramond, \& Ortolani 1989 \& Minniti et al. 1997).

The study by GHE01a used the 0\farcs1 internal\footnote{This refers to the
astrometric accuracy within a \cha\ image, while the accuracy of \cha's
astrometric reference frame is 0\farcs6 (Aldcroft et al. 2000)} positional
accuracy of \cha, and deep $UVI$ imaging of Gilliland et al. (2000; \hst\
program GO-8267) to confirm the two optical identifications of Verbunt \&
Hasinger (1998) and found \cha\ counterparts to V3 and AKO~9. A total of 13
optical identifications of CV candidates were found, the first detection of
a substantial population of such objects in 47~Tuc. A large population of
CVs in 47~Tuc was predicted by Di Stefano and Rappaport (1994), based on
tidal capture and Monte Carlo simulations. A later archival \hst\ study by
Ferraro et al. (2001a) also reported identifications of V1, V2, V3 and
AKO~9 with \cha\ sources, and an \hst\ study with the Space Telescope
Imaging Spectrograph (STIS) by Knigge et al. (2002) reported UV
counterparts for V1, V2, AKO~9 and W15 (a hard X-ray source identified by
GHE01a as having a faint blue optical counterpart). These follow-up studies
were limited by lack of sensitivity in the UV (Ferraro et al. 2001a) or by
the small $25''\times25''$ field of view (FoV) of the STIS camera (Knigge
et al. 2002; only 21 of the 103 GHE01a sources are found in the STIS FoV).

Other classes of X-ray binary have been optically identified in
47~Tuc. GHE01a reported 6 optical identifications of chromospherically {\it
active binaries}. In the galactic disk, active binaries include RS Canum
Venaticorum (RS~CVn) systems, typically consisting of a G- or K-type giant
or subgiant with a late-type MS or subgiant companion, and BY Draconis
(BY~Dra) systems, typically consisting of two late-type MS stars (Dempsey
et al. 1997). One of the active binaries identified by GHE01a shows a
likely X-ray flare and is identified with an optical variable discovered by
Edmonds et al. (1996).  A handful of active binaries have also been
observed in NGC~6397 (GHE01b) and NGC~6752 (Pooley et al. 2002a).

The sample of active binaries reported in this paper includes the X-ray
detection of several red stragglers (also known as sub-subgiants), objects
that are found just below the subgiant branch in globular and open
clusters. Six of these objects were found in 47~Tuc by Albrow et al. (2001;
hereafter AGB01), and others are known in NGC~6397 (GHE01b) and M67
(Mathieu et al. 2002).  Some red stragglers are in exotic binary systems
such as the subgiant secondary in the 47~Tuc CV AKO~9 and the optical
companion to the MSP 6397-A in NGC~6397 (Ferraro et al. 2001b; Orosz \& van
Kerkwijk 2003). However, they do not always contain degenerate objects,
since the recent study by Mathieu et al. (2002) of two red stragglers in
M67 identified one of them (S1113) as likely having a 1.3 \mdot\ subgiant
primary and a 0.9\mdot MS secondary. However, despite detailed photometric
(X-ray and optical), spectroscopic and proper-motion information, Mathieu
et al. (2002) have been unable to provide a secure explanation for these
objects.

Finally, three specialized papers have reported optical identifications of
neutron star binaries in 47~Tuc: a qLMXB (Edmonds et al. 2002a) and two
millisecond pulsars (MSPs) 47~Tuc~U (Edmonds et al. 2001) and 47~Tuc~W
(Edmonds et al. 2002b).  The 47~Tuc~U binary companion is a He~WD and the
companion to 47~Tuc~W is consistent with a heated main sequence star
showing large amplitude orbital variations.

In this two--paper series, we report: (1) full details of the CV and active
binary optical identifications first reported by GHE01a, (2) a large number
of new optical identifications using deeper photometric and time series
analysis of the extensive GO-8267 dataset, plus a general search for
optical companions to MSPs and qLMXBs and (3) analysis of archival F300W
and $V$ WFPC2 images obtained in \hst\ program GO-7503 (PI: G. Meylan). The
resulting set of optical IDs are by far the largest ever obtained for a
globular cluster. 

After a brief introduction to the data (\S~\ref{sect.obs}), the astrometry
for the optical counterparts will be described in \S~\ref{sect.ast} and the
optical photometry in \S~\ref{sect.phot}. The time series for these optical
counterparts will be presented in Edmonds et al. (2003; hereafter
Paper~II). A detailed analysis section will also be given in Paper~II,
including a study of the spatial distribution of the sources and their
X-ray to optical flux ratios.

\section{Observations}
\label{sect.obs}

\subsection{X-ray data}

The X-ray data used in this paper were obtained on 2000 March 16-17 and are
described in detail in GHE01a. The detection limit of about 3 counts for
these 72 ks, ACIS-I data corresponds to an X-ray luminosity, for an assumed
1~keV bremmstrahlung spectrum, and a 47~Tuc distance of 4.5 kpc (Heinke et
al. 2003), of 6$\times10^{29}$\ergs\ (in the 0.5--2.5 keV band). This
detection limit rises by factors of at least two or three within one core
radius ($r_c$) of the cluster center. For example only two (of 23) sources
within 0.5$r_c$ of the center have less than 12 counts.

We have retained the numbering used for the list of 108 sources given in
GHE01a.  This list was created by applying {\tt WAVDETECT} to the standard
level 2 event file, in energy band 0.5 to 4.5 keV, for a 2$'\times2.5'$
central field.  We have reprocessed the data using updated gain maps and
removing the standard pipeline randomization of event location within each
0\farcs5 pixel\footnote{See
http://asc.harvard.edu/ciao/threads/acispixrand/}, slightly improving the
resolution and astrometric quality of the \cha\ data.  We have used the new
coordinates for each source in the original list, except for twelve sources
that fell below our {\tt WAVDETECT} threshold with the new processing
because of crowding or low count levels; for these sources we use the
original coordinates. Visual examination of the original and reprocessed
images supports the reality of these 12 sources (as does the preliminary
analysis of new, much deeper \cha\ data).  In the cases where sources are
found in both the original and reprocessed lists, the positions typically
differ by only 0\farcs01-0\farcs05.

We have also included sources detected in the reprocessed data within the
$2'\times2\farcm5$ field, and beyond that field out to 4\farcm0 (extending
beyond any of the \hst\ data analyzed here).  These sources are assigned
names W109 to W193 (for concise reference to these sources in the text and
figures we have avoided giving their \cha\ convention source names). The
subset of these new sources that are within either the GO-8267 or GO-7503
FoVs are listed in Table~\ref{tab.extra-sources}.

\subsection{Optical data}

The GO-8267 \hst\ data (PI R. Gilliland) analyzed here is described in
detail in Gilliland et al. (2000) and AGB01. This program involved an
extensive set of 160~s WFPC2 exposures in F555W (`$V$'; 636 images) and
F814W ('$I$'; 653 images) obtained over 8.3 days in 1999 July. The exposure
times for the $V$ and $I$ data were designed so that stars about half a
magnitude brighter than the MS turn-off (MSTO) are saturated. In $V$ this
resulted in saturation for $\sim$1.0\% of the pixels on the least crowded
chip (WF3), and about 2\% of pixels on the most crowded chip (WF2). In $I$
the corresponding numbers are 1.5\% and 2.8\%.  A limited number (28) of
generally longer exposures in F336W (`$U$') were also obtained, with
negligible saturation levels (0.0002-0.001\% of pixels). Sub-pixel
dithering enabled deep, oversampled (by a factor of 4) images in $U$, $V$
and $I$ to be produced. The magnitude limits reached for stars on the MS
are $U\sim$23.5, $V\sim$23.5 (PC1 chip) to $V\sim$25.0 (WF3), and
$I\sim$21.5 (PC) to $I\sim$22.5 (WF3). Crowding and enhanced background are
the limiting factors in $V$ and $I$.

The other optical dataset analyzed in detail is the program GO-7503 (PI
G. Meylan) involving a short set of dithered F300W and $V$ exposures with
WFPC2 taken over six orbits about 110 days after GO-8267. The oversampled
F300W image has comparable depth to the GO-8267 $U$ image. The GO-7503 $V$
image combines only 18 short (20~s) exposures and therefore has much lower
signal-to-noise than the GO-8267 $V$ image, but the effective limiting
depth (in the CMD) differs by $\lesssim$1~mag because of the effects of
crowding in the GO-8267 data.  Two earlier epochs of F300W data obtained as
part of programs GO-5912 and GO-6467 are also briefly used in this paper in
testing for long-term variability.

\begin{figure}
%\vspace*{0.5cm}
%\hspace*{1.5cm}
%\epsfig{file=/data/edmonds/chandra/47tuc/paper-chandra-fov.eps,width=15.5cm}
\plotone{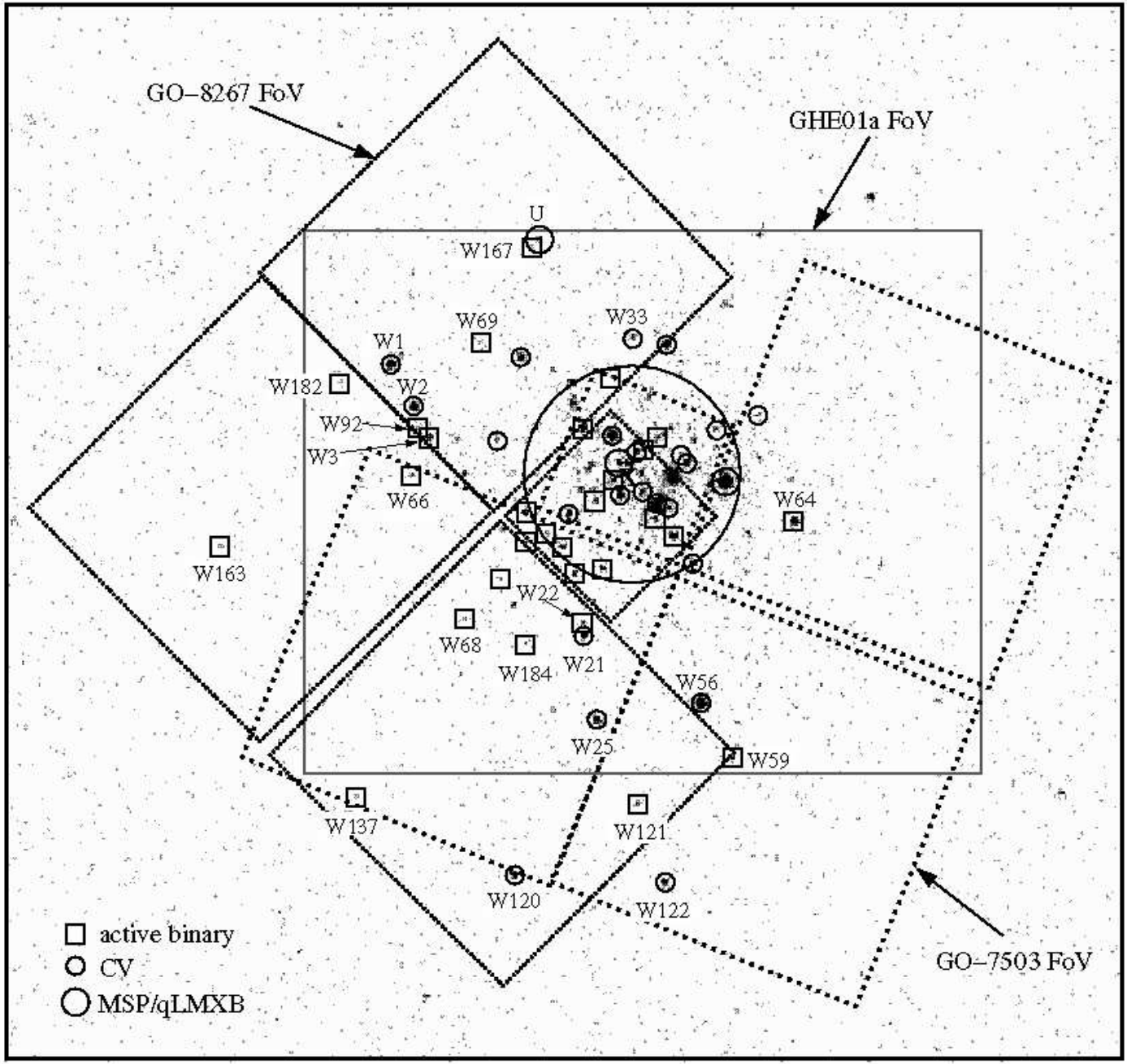}
%\vspace*{0.2cm}
\caption{A \cha\ image of 47~Tuc with the WFPC2 FoVs of the GO-8267 and
GO-7503 datasets overplotted, and the field analyzed by GHE01a shown by the
large rectangle. The approximate cluster center is shown by the `X' and the
24$''$ core radius is shown by the large circle. X-ray sources with optical
identifications are labeled with small circles for CVs, larger circles for
MSPs or qLMXBs and squares for active binaries (in each case the \hst\
positions, shifted to the \cha\ frame, are used to plot the optical
IDs). Marginal optical counterparts are not shown. Sources near the cluster
center are labeled in Fig.~\ref{fig.chandra-close}. Note the almost
complete lack of active binaries discovered in the regions observed only by
the GO-7503 program, because of the lack of high quality time series.}
%\vspace*{0.5cm}
\label{fig.chandra-fov}
\end{figure}

\begin{figure}
%\vspace*{0.5cm}
%\hspace*{-0.3cm}
%\epsfig{file=/data/edmonds/chandra/47tuc/paper-close-fov.eps,width=8.5cm}
\epsscale{0.65}
\plotone{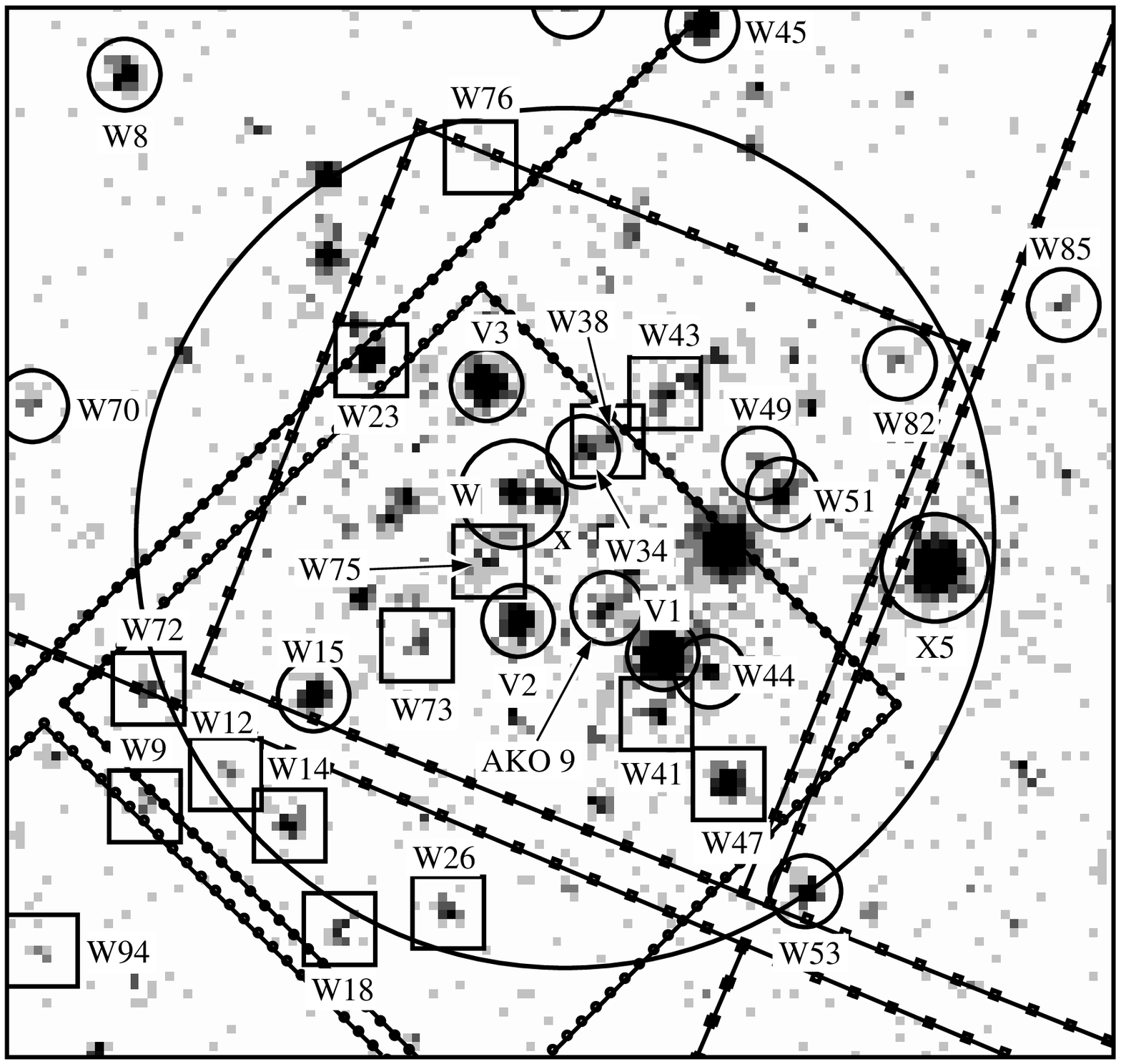}
%\vspace*{0.4cm}
\caption{A close-up of the central region of Fig.~\ref{fig.chandra-fov},
with use of the same labeling. Again, the \hst\ positions, globally shifted
to the \cha\ frame, are used for plotting the positions of the optical
IDs.}
%\vspace*{0.5cm}
\label{fig.chandra-close}
\end{figure}

The FoVs of the GO-8267 and GO-7503 datasets are shown in
Fig.~\ref{fig.chandra-fov} superimposed on the \cha\ image. The GO-7503
program was obtained at a roll angle that differs significantly from the
GO-8267 program, and hence the two FoVs are complementary. Also shown is
the 2$'\times$2.5$'$ field analyzed by GHE01a, the nominal center of the
cluster (the average of the positions quoted by Guhathakurta et al. 1992,
Calzetti et al. 1993 and de Marchi et al. 1996) and the 24$''$ core radius
(Howell, Guhathakurta, \& Gilliland 2000). A close-up of the center of the
cluster is shown in Fig.~\ref{fig.chandra-close}. A total of 78 \cha\
sources are found in the GO-8267 FoV and 84 in the GO-7503 FoV, with
significant overlap between these two lists especially near the center of
the cluster.  A complete list of these sources and their chip locations is
given in Table \ref{tab.other-names}.

\section{Astrometry}
\label{sect.ast}

We describe here the astrometric corrections used to shift the positions of
the \cha\ sources onto the \hst\ coordinate frame (random errors for the
WFPC2/\hst\ coordinates determined by {\tt STSDAS/METRIC} are assumed to be
negligible).  The optical identifications reported in GHE01a, Edmonds et al
(2001), Edmonds et al. (2002a), Edmonds et al. (2002b), and those to be
given here, were discovered using an iterative procedure, beginning with
the X-ray identifications of the first three CV candidates to be discovered
in 47~Tuc, V1 (Paresce, de Marchi, \& Ferraro 1992), V2 (Paresce \& de
Marchi 1994) and V3 (Shara et al. 1996).  The proposed astrometric matches
of Verbunt and Hasinger (1998) between V1 and X9 (corresponding to \cha\
source W42) and V2 and X19 (\cha\ source W30) were adopted. Then, after
applying the required offsets to the \cha\ positions in Right Ascension
(RA) and Declination (Dec), V3 was identified as a likely optical
counterpart to X10 (W27). The rms values of the residuals after subtracting
the 3 \hst\ positions from the 3 corrected \cha\ positions were determined
to be 0\farcs036 in RA and 0\farcs047 in Dec. Since we show that V1, V2 and
V3 are both blue (see \S~\ref{sect.phot}) and variable (see Paper~II), and
the chance of these matches being a coincidence is extremely low (see
\S~\ref{sect.phot}), these stars are confirmed as the optical counterparts
of X9, X19 and X10 respectively.

Based on this determination of the \cha/\hst\ bore-site, we searched the
GO-8267 data for optical variables and stars with colors outside those of
the main sequence or giant branch.  Any such objects lying within
3-$\sigma$ of the positions of the \cha\ sources, using the positional
errors given by {\tt WAVDETECT}, were identified as optical counterparts.
Using the GO-8267 data we found 10 or more candidate optical counterparts
for \cha\ sources on each of the PC1, WF2 and WF4 chips. We then tested for
linear correlations between the astrometric residuals and their positions,
with appropriate weighting of each source using the errors estimated by
{\tt WAVDETECT}, and then removed this correlation.  For both the PC1 and
WF2 chips we found a significant correlation between the RA and RA
residual, with constants of proportionality ($a$) of 0.0053$\pm$0.0011 for
the PC1 and 0.0024$\pm$0.00049 for WF2. For Dec no such correlations were
found ($a$= 0.0009$\pm$0.0010 for PC1, and $a$= 0.0003$\pm$0.0015 for WF2).
These results imply that systematic errors (of unknown origin) are present
in the relative \hst-\cha\ positions, but even for stars near the extreme
edges of the WF chips the astrometric corrections are small, no greater
than $\sim$0\farcs15. For the WF3 chip there were only 3 \cha\ sources with
optical counterparts, and here only the bore-site correction was
calculated. For the WF4 chip we found a marginally significant correlation
in RA, but with the opposite sign. Here, the astrometric solution is
dominated by the only three sources (W21, W25 and W120) with more than 50
counts, and they are separated in RA by only $\sim$18$''$, so the
systematics are less accurately constrained. In Dec we found a marginal
systematic error for WF4 ($a$= 0.0026$\pm$0.0011).

For the 5 sources with more than 50 counts on the PC1 chip, the rms
residuals in RA and Dec were determined to be 0\farcs053 and 0\farcs035
respectively before the linear corrections and 0\farcs014 and 0\farcs035
after the corrections. For WF2 the RA residual was reduced from 0\farcs068
to 0\farcs050 for the 5 sources with $>50$ counts (no significant change in
Dec values) and for WF4 no significant improvement was found in RA but the
Dec rms decreased from 0\farcs075 to 0\farcs019 for the 3 sources with
$>50$ counts.

For comparison, we have performed a similar astrometric solution for the 6
most isolated X-ray counterparts to the MSPs. Here, we found the same
apparent systematic error in the \cha\ positions for RA, relative to the
radio positions, with $a = 0.0059 \pm 0.0021$, a value consistent with that
found for the \cha/\hst\ comparison (see also Edmonds et al. 2001). This
confirms that there are small systematic errors in the \cha\
positions. Deeper \cha\ observations (recently obtained; see Paper~II) will
give much better source locations for the MSPs (and a more complete source
list), and will allow further study of these subtle astrometric errors.

We included all of the optical IDs in a global astrometric solution, but
here the brightest sources on the PC1 chip dominated the astrometry.
Therefore, because different constants of proportionality were appropriate
for each \hst\ chip, and because we found small ($\sim0\farcs05-0\farcs1$)
offsets between chips, this global solution gave larger errors than for the
chip-by-chip solutions. Therefore, this procedure was discarded and the
separate chip solutions were used.

\begin{figure}
%\vspace*{-2.5cm}
%\hspace*{0.5cm}
%\epsfig{file=../fchart/paper-fchart.eps,width=16.5cm}
\epsscale{1.0}
\plotone{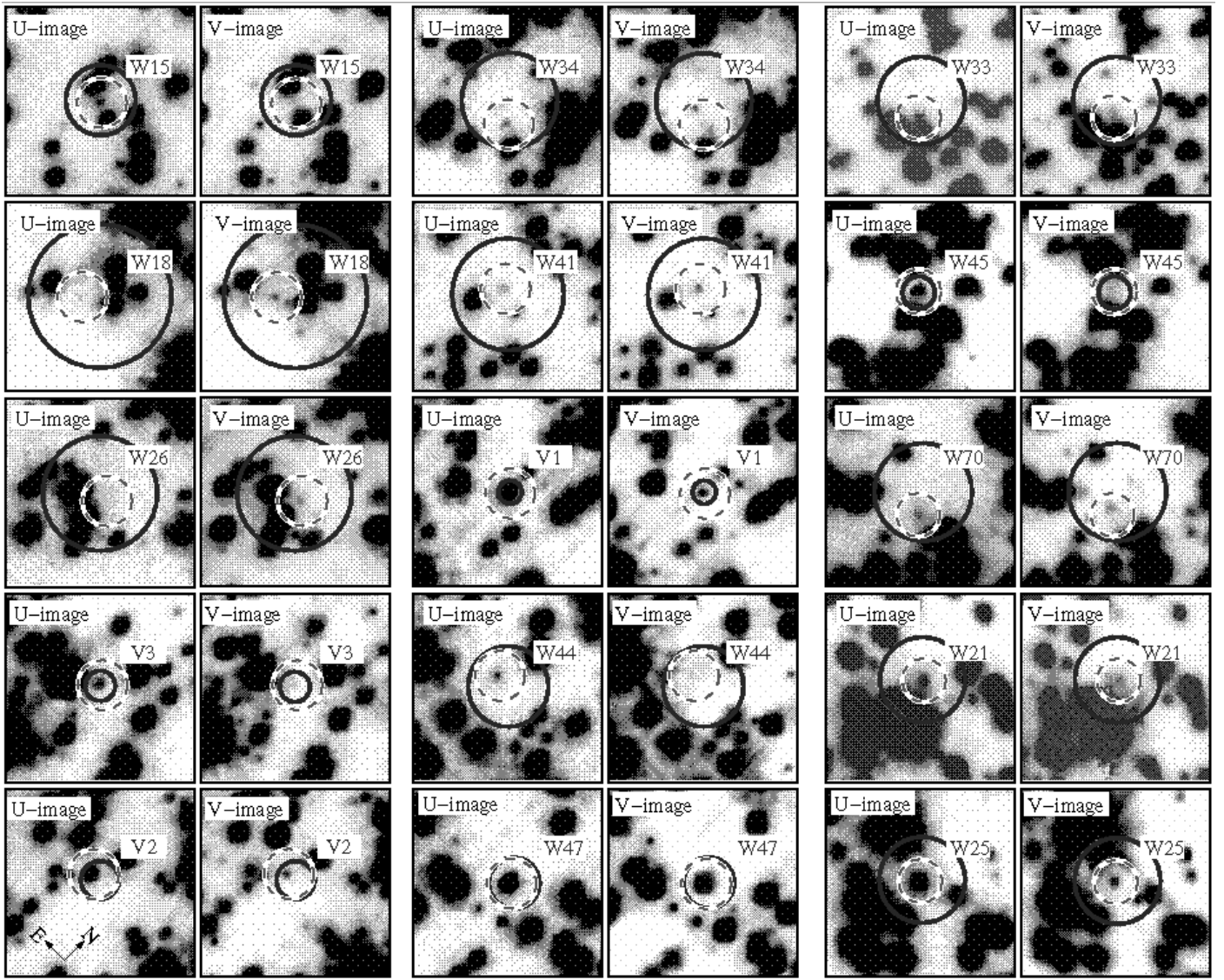}
%\vspace*{0.2cm}
\caption{A sample of \hst\ finding charts for 15 optical identifications
from the GO-8267 dataset. Optical IDs for CV candidates in the PC1 chip
(W15\opt, V3, V2, W34\opt, V1, W44\opt), the WF2 chip (W33\opt, W45\opt,
W70\opt) and the WF4 chip (W21\opt, W25\opt) are shown, plus optical IDs
for active binary candidates in the PC1 chip (W18\opt, W26\opt, W41\opt,
W47\opt). W34\opt\ may also be an MSP. In each case the deep, oversampled
images in F336W ('U') and F555W ('V') are shown. The solid circles show the
4-$\sigma$ errors (the \cha\ error and the systematic errors for the
\cha-\hst\ transformation added in quadrature) and the dashed circles show
the candidate optical IDs (radius = 0\farcs2 for the PC1 chip and 0\farcs3
for the WF chips). The CV candidates are all relatively bright in the
$U$-band, and the active binary candidates are mostly brighter in the
$V$-band.}
%\vspace*{0.5cm}
\label{fig.fchart}
\end{figure}

Similar analysis was applied to the GO-7503 data and will not be described
in detail here. Tables \ref{tab.8267} and \ref{tab.7503} show the RA and
Dec offsets for all of our candidate optical counterparts (W38, W92 and W94
positions were based on the randomized data used in GHE01a), including
several marginal counterparts. Alternative designations for the optical IDs
are given in Table \ref{tab.other-names}.  All of the counterparts are
within 3-$\sigma$ of the quoted X-ray source (where the $\sigma$ values are
the {\tt WAVDETECT} errors added to the systematic errors from our linear
fits in quadrature), except for the marginal W71 counterpart at
$\sim3.4\sigma$, which we have included because the star is both blue and
variable. Further discussion of the astrometry for this star will be given
in Paper~II. Finding charts for a sample of the optical
counterparts are given in Fig.~\ref{fig.fchart}.

\begin{figure}
%\vspace*{0.5cm}
%\hspace*{-0.2cm}
%\epsfig{file=/data/edmonds/chandra/47tuc/paper-marg-fchart.eps,width=8.5cm}
\epsscale{0.65}
\plotone{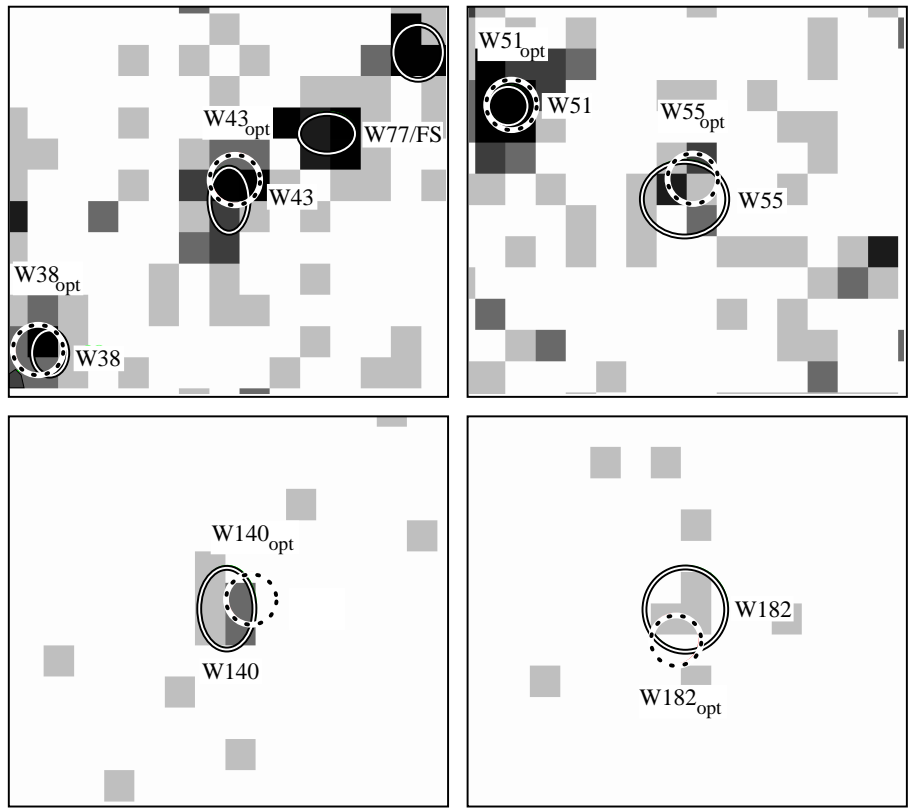}
%\vspace*{0.2cm}
\caption{\cha\ images for 4 optical counterparts that are relatively
marginal, based on astrometry. The solid lines show the \cha\ 5-$\sigma$
error ellipses and the dashed lines show the position of the possible
optical counterparts. The ID for W43 is a likely red straggler and
therefore is a reasonably secure counterpart (the presence of at least two
unresolved X-ray sources appears to have skewed the position of the X-ray
source). The source W55 also shows evidence of crowding, but W140 and W182
both appear to be relatively isolated sources.}
%\vspace*{0.5cm}
\label{fig.marg-fchart}
\end{figure}

\begin{figure}
%\vspace*{0.5cm}
%\hspace*{-0.2cm}
%\epsfig{file=/data/edmonds/chandra/47tuc/paper-bluevar-fchart.eps,width=8.5cm}
\epsscale{0.5}
\plotone{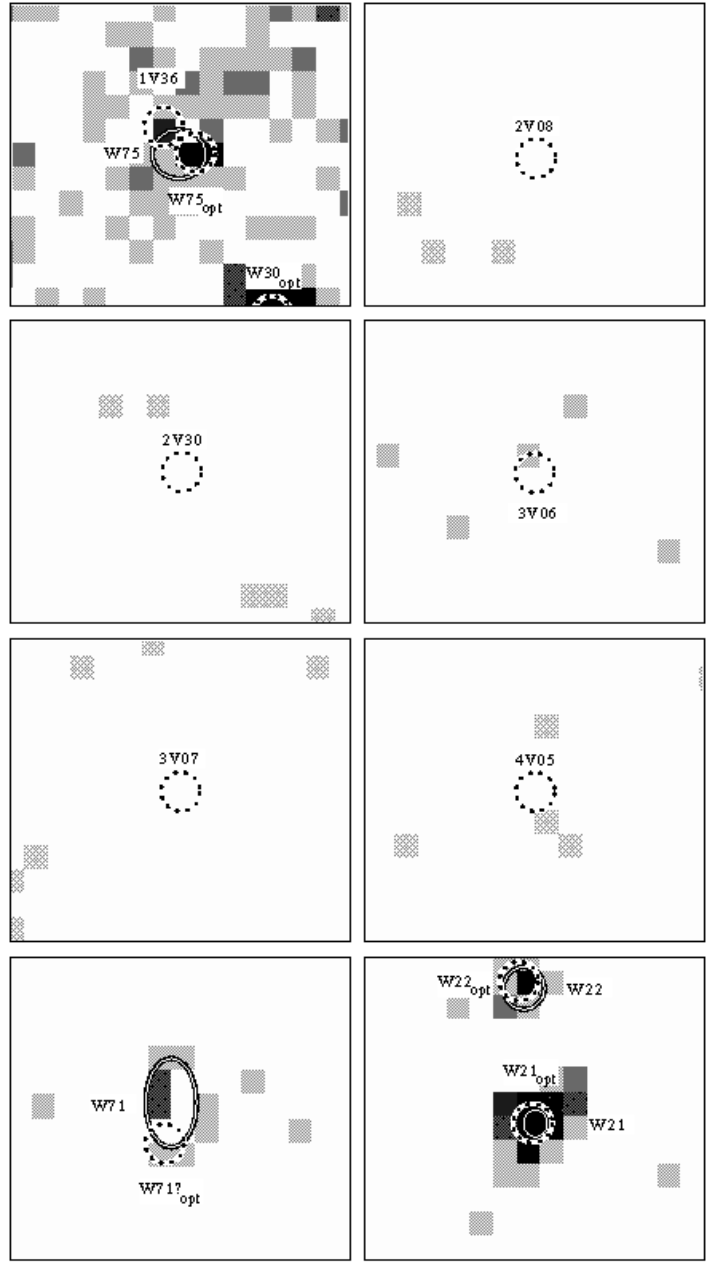}
%\vspace*{0.2cm}
\caption{\cha\ image with overplotted circles for the blue variables 1V36,
2V08, 2V30, 3V06, 3V07, 4V05 from AGB01 (the notation [n]V[m] is used where
[n] gives the WFPC2 chip number and [m] gives the sequential variable
number), plus W71 with its marginal X-ray ID. The same labeling as used in
Fig.~\ref{fig.marg-fchart} is adopted. Source W21 is shown because it has
optical colors that are very similar to most of the other blue
variables. The blue variable 1V36 may correspond to a faint X-ray source,
but the other 5 blue variables clearly do not lie near detectable X-ray
sources. Error ellipses for the \cha\ sources found with {\tt WAVDETECT}
are also shown.}
%\vspace*{0.5cm}
\label{fig.possblue-fchart}
\end{figure}

Besides the possible counterpart for W71, only the W43, W55, W75, W140 and
W182 counterparts are at $>2.0\sigma$ (W73\opt\ is at 2.04$\sigma$ in the
GO-7503 data but is at 1.89$\sigma$ in GO-8267).  Finding charts showing
the \cha\ image (0.5-8 keV) with these possible IDs overplotted are shown
in Fig.~\ref{fig.marg-fchart} (W43, W55, W140 and W182) and
Fig.~\ref{fig.possblue-fchart} (the possible IDs for W71 and W75). The
\cha\ errors may have been underestimated for W43, W55 and W75 because they
are close to cluster center and are relatively crowded. For example, the
source W43 appears to be blended with a faint nearby source not detected
with {\tt WAVDETECT}, and W75 may be embedded in weak diffuse emission
(possibly from a large number of weak sources).  For W182 the presence of
only 3 \cha\ sources with likely optical counterparts on the WF3 chip has
limited the accuracy of our astrometry. The likelihood of chance
coincidences are described in \S~\ref{sect.phot}.

\section{Photometry}
\label{sect.phot}

The photometry analysis of the GO-8267 dataset used standard IRAF and
DAOPHOT tools and customized software written by PDE.  Since we required
photometry for faint red stars (active binaries) as well as blue objects
(CVs and MSPs), separate star lists were created for each filter and the
DAOPHOT program {\tt ALLSTAR} was run on each list.  Stars were included in
a color magnitude diagram (CMD) if they were detected in either all 3
filters or in $U$ and $V$ or $V$ and $I$ (the maximum matching distance
between filters was 2 oversampled pixels).  Objects with poor PSF fits and
those with large amounts of light contamination from neighboring stars were
removed. Despite this, a large number of artifacts remained in the $V$ vs
$V-I$ CMD, caused by the effects of structure in the PSFs of bright (often
saturated) stars in the very deep $V$ and $I$ images.  More sophisticated
techniques are required to improve the star lists and eliminate these
artifacts, however these methods are unnecessary for this paper. Since the
$U$ image is affected less by artifacts (it is not as deep and the giant
stars are not nearly as badly saturated), the $U$ vs $U-V$ CMD is much
cleaner than the $V$ vs $V-I$ CMD. Since most of our X-ray counterparts are
detected in $U$, and the exceptions are variables identified in $V$ and $I$
difference images, our list of X-ray IDs is likely to be clean of
artifacts.

Calibration of the F336W, F555W and F814W magnitudes into $U$, $V$ and $I$
was performed using the zeropoints of AGB01 and the color corrections of
Holtzman et al. (1995). The resulting $U$ vs $U-V$ and $V$ vs $V-I$ CMDs
are shown in Figures \ref{fig.gill-pc1wf2} and \ref{fig.gill-wf3wf4}, and
close-ups of the region near the MSTO are shown in Fig.~\ref{fig.msto},
where the data for all 4 chips is superimposed. Numerals are used to denote
likely X-ray counterparts (using the numbering of GHE01a and our extended
numbering system introduced earlier).  Since several of the X-ray IDs are
faint and crowded (CVs and a few active binaries) or are partly saturated
(blue stragglers) we relaxed two of our CMD selection criteria (crowding
for the CV and active binary candidates and PSF-fit quality for the
saturated blue stragglers) to plot the fullest possible set of X-ray IDs on
this figure.  There were several CV candidates where the PSF-fit quality in
both the $V$ and $I$ bands exceeded our threshold and/or the position of
the fitted star was more than 2 pixels away from the $U$ band position
(W15\opt, W33\opt, W45\opt, W70\opt). Here, $V$ magnitude estimates were
performed by iteratively using ADDSTAR to subtract PSF models of the star
in the $V$ image until the residuals were minimized.  Larger errors were
appropriate here than for the overall sample of stars, particularly for
W33\opt\ and W70\opt\ where the $V$ counterpart to the object detected in
the $U$ band is difficult to see in the oversampled images. However, in
each case the stars are clearly blue by visual examination of the deep
images (see Fig.~\ref{fig.fchart}). No such fitting was attempted in $I$
where the crowding was even worse.

The analysis of the GO-7503 photometry (F300W and F555W) was carried out
using the same PSF-fitting that was used for the GO-8267 analysis. The only
difference in the analysis technique was that the deep oversampled images
were created using the suite of drizzle routines (Hook, Pirzkal, \&
Fruchter 1999) available in STSDAS (these are designed to handle cosmic-ray
removal with sparsely dithered data, unlike our customized software used
for the GO-8267 analysis). The standard procedure recommended in the
analysis guide was used, but with the two `SNR' and one `scale' parameters
in {\tt driz\_cr} set to small values (1.0, 3.0; 2.5) to ensure as many
cosmic rays were removed as possible. The disadvantage of this procedure is
that some bright stars contained cores that were falsely identified as
cosmic rays, but this did not affect our primary search for faint blue
stars.  The cosmic ray-cleaned F300W and F555W images for each epoch were
combined into doubly oversampled images using the Drizzle routine.  Because
the GO-7503 data was mainly useful for detecting blue stars (CV or MSP
candidates) that are brighter in the F300W image than in the F555W image,
the deep F300W image was used to create a master star list.  Using this
star list, PSF-fitting was applied to both the F300W and F555W images using
DAOPHOT programs. After applying a plate solution (using bright, relatively
isolated stars used to build the PSF model) to transform the F555W
coordinate frame to that of the F300W frame, stars $<2$ oversampled pixels
apart in the two filters were considered detected in each. Extra filtering
of the results was performed by eliminating stars with chi-squared values
$>3$ and with `sharpness' values $> 1.2$ in the {\tt ALLSTAR} output. Most
of the stars with high chi-squared values were bright stars with central
pixels that were mistaken for cosmic rays and were therefore truncated,
distorting the PSF in the summed image. Since CVs are usually much fainter
than the MSTO, the loss of a few bright MS stars and some giants did not
affect our results.  Stars with large sharpness values were generally PSF
artifacts near bright stars.  Figure \ref{fig.meycmds} shows the CMDs
resulting from this analysis, with labeling of likely or possible optical
IDs to X-ray sources.

We also included stars detected in F300W but not in F555W.  To avoid
including large numbers of artifacts (mainly false stars detected in the
PSF wings of bright stars) we set an upper limit on the total counts in a
2-pixel radius aperture in the F555W image. This limit did not seriously
compromise our efforts to detect faint blue stars, since no candidate CVs
were lost in this way. In cases where IDs are detected in F300W but not
F555W the red limits are shown with arrows (the limits in F555W were
determined from the 3$\sigma$ limit in the aperture counts).  Other such
red limits are not shown in Fig.~\ref{fig.meycmds} to avoid confusion.

\subsection{Cataclysmic variables}
\label{sect.cvphot}

Thirteen CV candidates in 47~Tuc have already been presented in GHE01a: V1
(W42), V2 (W30), V3 (W27), W1\opt, W2\opt, W8\opt, W15\opt, W21\opt,
W25\opt, W36\opt\ (AKO~9), W44\opt, W45\opt\ and W56\opt. These are blue
stars astrometrically matched ($<3\sigma$) to \cha\ sources, and all but
one of them are found in the GO-8267 FoV, and are plotted in
Fig.~\ref{fig.gill-pc1wf2}. The exception (W56\opt) was initially found in
analysis of archival F300W data and is confirmed by the GO-7503 analysis
(see below). Seven extra CV candidates (W33\opt, W34\opt, W35\opt, W70\opt,
W120\opt\ and W140\opt, plus the marginal ID for W71) were found by closer
examination of the GO-8267 images. Only two of these (W120\opt\ and
W140\opt) were outside the 2$'\times 2.5'$ field analyzed by GHE01a.

The CV candidates in the GO-8267 data are all detected as clearly blue in
the $U$ vs $U-V$ CMDs, with the exception of AKO~9 where the light in the
$U$ band has a significant contribution from the subgiant secondary. This
star is redder in the $V$ vs $V-I$ CMD, where the subgiant dominates.  A
few of the CVs have blue $V-I$ colors (e.g. W1\opt, W21\opt\ and V1), but
most of them are found close to the MS ridge-line (although, as noted
above, several of the optical counterparts were too faint or crowded to be
detected by PSF-fitting in the $I$ band).  A similar result was found by
Cool et al. (1998) for the CVs in NGC~6397 (we encourage detailed fitting
to the broadband photometry of the CVs in both 47~Tuc and NGC~6397
to help constrain the properties of the accretion disk and compare with
field CVs).  A candidate ID for W35 is unfortunately affected by
diffraction spikes in $V$ and $I$ (a saturated star with $V$=14.0 is within
0\farcs34 or 3.0$\sigma$ of the nominal W35 position), but it does appear
to be blue from visual examination.

In each of the GO-8267 CMDs the hydrogen WD models of Bergeron, Wesemael,
\& Beauchamp(1995) and the He~WD models of Serenelli et al. (2001) are
shown.  Fig.~\ref{fig.gill-wf3wf4} also shows CVs from the field surveys of
Zwitter \& Munari (1995), Zwitter \& Munari (1996), and Munari \& Zwitter
(1998) and the 3 brightest He~WDs and the 4 brightest CVs in NGC~6397 (Cool
et al. 1998). Note the expected clear separation between the CVs (both
field and cluster) and the cluster MS stars. Several of the 47~Tuc CVs have
much redder $V-I$ colors than the field and the NGC~6397 CVs.

Of the GO-8267 data CV candidates lying also in the GO-7503 FoV (V1, V2,
V3, W15\opt, W21\opt, W25\opt, W34\opt, AKO~9, W44\opt, W120\opt\ and the
marginal W71 ID) all are shown as blue stars in Fig.~\ref{fig.meycmds}
except for V3 which fails both our chi-squared and sharpness tests
(probably because large amplitude, short time-scale variability distorts
the PSF in the process of combining the dithered images) and W120\opt\ that
lies just at the outer edge of the FoV. However, both of these exceptions
are easily confirmed as blue stars by visual examination. A possible
counterpart to W31 is in the FoV of both HST datasets but is only detected
as a candidate blue star in the GO-7503 data. This is a marginal detection
requiring confirmation with other data, since the possible blue object is
close to a bright star, and is offset at the correct angle to be affected
by diffraction spike artifacts.  The possible CV identification for W35
also appears to be blue in the GO-7503 data, but the blue excess is clearly
not as dramatic as for the CV candidates discussed above
(Fig.~\ref{fig.meycmds}), and we continue to classify this ID as marginal.

There are 8 blue stars and CV candidates that fall just in the GO-7503 FoV
(W49\opt, W51\opt, W53\opt, W55\opt, W56\opt, W82\opt, W85\opt\ and
W122\opt). Only W56\opt\ and W122\opt\ were detected in both F300W and
F555W. As noted above, the positional discrepancy is relatively large for
W55\opt.

From examination of Figures \ref{fig.gill-pc1wf2}, \ref{fig.gill-wf3wf4}
and \ref{fig.meycmds} it is clear that there are significant numbers of
blue objects found in the two datasets analyzed here. There are also blue
stars that are detected in F300W but not F555W. To include both of these
classes of object we identify stars as blue if they are (1) more than
$3\sigma$ bluewards of the MS or (2) if they are detected in F300W but not
F555W and which have red limits that are also bluewards of the MS.  Based
on this analysis (for GO-7503), 181 blue stars were found on PC1, 256 on
WF2, 170 on WF3 and 158 on WF4.  These include a relatively small number of
CVs (and possibly a few undetected MSPs) combined with cluster WDs and MS
stars from the SMC, plus some artifacts. Given these large populations, we
have calculated the probability that our astrometric matches between \cha\
sources and blue stars are chance coincidences, based on the GO-7503 data
(this has comparable depth to the GO-8267 data). We have calculated the
number of blue stars (including stars with upper limits in F555W that
appear to be blue) per unit area as a function of radial distance from the
cluster center.  We applied an evenly spaced grid (with individual grid
elements of 1\farcs6) across the WFPC2 FoV and corrected for the incomplete
radial coverage. To include the effects of crowding we deleted grid points
with aperture counts in F555W greater than the threshold mentioned
earlier. The resulting radial density function represents an upper limit to
that of the true blue star population above our detection threshold because
(a) PSF artifacts likely remain in our star list despite the quality
controls applied to the PSF fitting results, and (b) elimination of grid
points for counts above our threshold is only appropriate for stars
detected only in F300W.  Applying this function at the appropriate radial
distance for each CV candidate and multiplying by $\pi\times $offset$^2$
(where `offset' is the \hst/\cha\ offset for each counterpart) to scale by
area, we found probabilities of chance coincidences as given in the 6th
column of tables \ref{tab.8267} and \ref{tab.7503}. Note that only for W55,
W71 and W140 are the probabilities $>$4\%.

\subsubsection{Previous Observations}
\label{sect.prevobs}

As noted in \S~\ref{sect.intro} and \S~\ref{sect.ast}, and as shown in
Table \ref{tab.other-names}, several of the CVs presented here
were discovered before GHE01a: V1 (Paresce, de Marchi, \& Ferraro 1992), V2
(Paresce \& de Marchi 1994), V3 (Shara et al. 1996) and AKO~9. The latter
object is variable \#11 from Edmonds et al. (1996), later identified with
AKO~9 (Auriere, Koch-Miramond, \& Ortolani 1989) by Minniti et al. (1997).

All 4 of the above objects have been identified in the \hst\ survey of
Ferraro et al. (2001a), with nominal angular separations between the \cha\
sources and the blue stars of 0\farcs0--0\farcs1. We believe that the
majority of the other possible associations between \cha\ sources and blue
stars given in Table 5 of Ferraro et al. (2001a) are chance coincidences,
and we note the appropriately tentative support given by Ferraro et
al. (2001a) regarding the reality of these IDs. Except for the clearly
identified objects described above, 17 of the 25 remaining suggested IDs
have angular separations $>$0\farcs9, and these typically have astrometric
errors that are $>5\sigma$ (when using the \cha\ derived errors) and in
some cases are $>10\sigma$. These offsets are much larger than those given
in Tables \ref{tab.8267} and \ref{tab.7503}, except for the few marginal
candidates, and scaling from our numbers in Tables \ref{tab.8267} and
\ref{tab.7503} the chances of coincidental matches with blue stars are
considerable. 

The objects with suggested IDs at large separations often include X-ray
sources with more reasonable IDs presented here in Tables \ref{tab.8267}
and \ref{tab.7503}. For example, W15, W34 and W44 all have variable
counterparts lying much closer to the X-ray position. The suggested blue
straggler IDs for W31 and W37 have offsets of 0\farcs4 and 0\farcs5, both
worse than 3-$\sigma$ astrometrically (W31 has a marginal detection that
lies considerably closer to the \cha\ position). Also, W51 and W75 both
have more likely IDs than those presented in Table 5 of Ferraro et
al. (2001a; see below for further comments on the suggested match to
W75). The possible ID for W98 appears to be only marginally blue in the
$m_{F218W}$ vs $m_{F218W}-m_{F439W}$ CMD and lies at an unreasonably large
angular separation of 0\farcs7. We find no evidence of a blue star in our
deep, oversampled images, although crowding from giants stars is a problem
for this source. The remaining two suggested IDs are those for W54 and W80
at separations of 0\farcs1 and 0\farcs2. We believe that the bright, red ID
for W54 may be an RS~CVn, but we see no evidence for a blue object (or a
star with long-term variability) near W80.

The \hst/STIS survey of Knigge et al. (2002) has identified the UV
counterparts to V1, V2, AKO~9 and W15\opt. None of the CV candidates of
Knigge et al. (2002) are close to either W31 or W35, and no obvious star is
nearby in the UV image apart from a bright blue straggler offset slightly
from W31.  This suggests that these two objects may not be CVs, although
confirmation of this is needed by direct analysis of the UV image, since
nearby bright stars may be preventing the detection of faint CVs in the
$U$-band image (see also \S~\ref{sect.unid}). Several X-ray sources
mentioned as having possible IDs from Ferraro et al. (2001a) fall in the
Knigge et al. (2002) FoV. Of these, W31, W37 and W98 are not suggested by
Knigge et al. (2002) as CV candidates. The possible ID for W75 (Knigge et
al. 2002) will be discussed in Paper~II.  The general lack of detection of
UV counterparts to X-ray sources by Knigge et al. (2002) will be discussed
in \S~\ref{sect.abs} and \S~\ref{sect.unid}.

\subsection{Blue Variables}
\label{sect.bluevarphot}

In Figures \ref{fig.gill-pc1wf2} and \ref{fig.gill-wf3wf4} we also show
blue variable stars that were discussed by AGB01 as possible CV candidates
(PC1-V36, WF2-V08, WF2-V30, WF3-V06, WF3-V07, WF4-V05 and WF4-V26,
abbreviated as 1V36, 2V08, 2V30, 3V06, 3V07, 4V05 and 4V26), but were not
identified with X-ray sources by GHE01a. We have carefully examined these
{\it blue variables} in the deep, oversampled \hst\ images, and confirmed
that all of them (except 4V26) have bluer colors than most other cluster
stars, and are unaffected by artifacts (Table~\ref{tab.bluevar} lists the
colors, positions and periods for these objects).  Two blue, apparently
variable objects from AGB01 were found to be possible artifacts: PC1-V52
has an irregular light curve but is not obviously blue and is found near a
diffraction spike, and the apparent blue color and irregular variability of
WF4-V16 appears to be caused by a bad pixel.

Most of the blue variables (4 out of 6; 3V06 does not have an $I$
magnitude) are found in a region of the color-color plot where no field CVs
are found, at $0\lesssim U-V\lesssim0.4$, $0\lesssim V-I\lesssim0.3$
(Fig.~\ref{fig.close-colcol} shows a close-up of the color-color plot). Of
the 5 objects in this area only W21\opt\ is clearly an X-ray source, so it
is the only strong CV candidate. The other two blue variables (2V08 and
2V30) have significantly different colors, suggesting that they may
represent a different class of object. Figure~\ref{fig.possblue-fchart}
shows the \cha\ image (0.2--8 keV) with the blue variables (plus W21\opt)
overplotted as circles.  Further discussion of these blue variables will be
given in Paper~II.

\subsection{Active Binaries}
\label{sect.abs}

As shown by Tables \ref{tab.8267} and \ref{tab.7503} and Figures
\ref{fig.gill-pc1wf2} and \ref{fig.gill-wf3wf4}, a large number of the
\cha\ sources are astrometrically matched to optical variables with the
colors of MS stars or subgiants (with no evidence for blue components).  We
believe the majority of these optical variables are active binaries
(BY~Dras or RS~CVns), though a small fraction could be other possibilities
such as MS stars with MSP companions (see Paper~II). The clearest example
of an active binary is the \cha\ source W47 identified by GHE01a with
optical variable \#8 from Edmonds et al. (1996) and PC1-V08 from AGB01. The
source W47 shows clear variability in the \cha\ dataset (see Fig.~5 of
GHE01a), a likely flare signature.  Five other active binaries were
reported by GHE01a: W14\opt, W18\opt, W41\opt, W68\opt\ and W43\opt. The
first three of these were also identified by AGB01 (and are also found in
the GO-7503 field of view), the 4th is found in both the GO-8267 and
GO-7503 datasets but fell just below the variability detection thresholds
used in AGB01 (see below) and the 5th is a likely red straggler in the
GO-7503 data (see Fig.~\ref{fig.meycmds}). Red stragglers were defined by
AGB01 as variable stars lying inside the following box in the CMD: 0.8 $<
V-I <$1.5; 17.25$< V <$17.75 (see Fig.~\ref{fig.msto}), but obvious
candidates are visible in CMDs using different broadband colors (see
Fig.~\ref{fig.meycmds}).

The 6 active binaries of GHE01a were selected using the conservative
criterion for an active binary that the X-ray source be variable. This
criterion is conservative because faint (5--20 ct) sources must show
variability that is intrinsically extreme to be considered statistically
significant. Two other active binary candidates are matched to variable
X-ray sources (W92\opt\ and W94\opt), and an additional 20 optical
variables and active binary candidates are matched to statistically
non-variable X-ray sources.

Most of this large sample of active binary candidates have been
independently discussed by AGB01 and include variables classified as
eclipsing binaries (W12\opt, W92\opt, W137\opt, \& W182\opt), W~UMa
binaries (W41\opt, W47\opt, \& W163\opt), a non-eclipsing contact and
semi-detached binary (W66\opt), red straggler variables (W3\opt\ \&
W72\opt) and `BY~Dra' variables\footnote{AGB01 use the term BY~Dra for low
amplitude variables on or close to the MS that do not appear to be elipsing
or contact systems, but we use this term in a more specialized sense to
refer to systems that are both X-ray sources and optical variables
containing two MS stars} (W9\opt, W14\opt, W18\opt, W69\opt, W73\opt,
W75\opt, \& W76\opt). Table \ref{tab.other-names} gives the AGB01
designations for each of these objects.  Only the association between \cha\
X-ray sources and some of the red stragglers was noted by AGB01.

A number of other optical variables in the GO-8267 FoV (not reported by
AGB01) were discovered by searching carefully at the positions of X-ray
sources. The new detections were for \cha\ sources W22, W23, W26, W38, W59,
W94, W121, W167, \& W184. The ID for W59 was discovered with the
single--pixel search method, where we searched for statistically
significant periodicities in the time series of all individual pixels in
the GO-8267 data-set (a valuable technique for detecting variable stars
that are near artifacts such as saturation trails). This variable is near
the edge of WF4 and is close to a bright neighbor. The ID for W167 (a blue
straggler) was initially detected as a variable but was then dropped from
the AGB01 list because it is saturated. The other variables were not
detected in the original AGB01 source list due to the faintness of the
stars and crowding. The time series for these new variables will be
discussed in Paper~II.

Several other X-ray sources have possible identifications with stars that
are either potential red stragglers or that have marginally significant
variations in the optical (see Fig.~\ref{fig.msto}).  For continuity these
results are quoted here, although most of the time series results are
presented in Paper~II.  The possible ID for W4 is identified with a
non-variable star near the blue side of the box in the CMD defining red
stragglers, but without the detection of variability we classify this
possible red straggler as a marginal candidate.  The source W37 is a
variable X-ray source within 2$\sigma$ of two bright ($V$=17.4 and 17.1) MS
stars. Either one of these stars could be an active binary but because
variability is not observed we do not label them as such. The ID for W68,
one of the original GHE01a active binaries, is a strongly variable ($>$99\%
significance) X-ray source lying close to a star showing no obvious
periodic variations in $V$ but with a likely 1.12 day period in $I$. The
false-alarm probability (FAP; Horne \& Baliunas 1986) equals $5.8 \times
10^{-4}$ at the power spectrum peak at 0.56 d. We consider this to be a
reasonably secure detection. Finally, the marginal counterpart for W93 has
a period of $0.1122\pm0.002$ days and an amplitude of $0.0016\pm0.0003$
(5.1$\sigma$). If real, the orbital period is probably twice as long.

We now consider if any of these astrometric matches are likely to be chance
coincidences. We calculated the spatial density of variables discovered by
AGB01 for each of the 4 WFPC2 chips, scaling these densities by appropriate
factors to account for variables not detected by AGB01 but subsequently
found in deeper searches around the X-ray positions. Using this technique,
we found that the largest chance probability for an active binary candidate
is for W75\opt\ (1.1\%) and all of the other active binaries have chance
probabilities $<$0.67\% (see the 6th column of Tables \ref{tab.8267} and
\ref{tab.7503}).  Given these small probabilities we expect that the number
of chance coincidences is negligible.

If the active binaries are double MS binaries then they should be found
either on the MS or above it by $\lesssim$ 0.75 mag. We tested this
hypothesis by measuring the distance of each of the active binaries above
the MS ridge-line and comparing the distribution of these offsets with
those found for the AGB01 binaries and the general stellar population. We
performed iterative fits to the MS ridge-lines in both the $U$ vs $U-V$ and
$V$ vs $V-I$ CMDs (after removing 3-$\sigma$ outliers) and then measured
the vertical offsets ($\Delta$ mag at fixed color) of each star from these
fits. To avoid regions near the MSTO where the MS ridge-line is vertical
(or close to it) we restricted our studies to regions with $U>18.4$ and
$V>18$, which imposed blue limits in $U-V$ (0.53) and $V-I$ (0.72). We only
included X-ray IDs that survived the PSF quality and crowding tests
presented in \S~\ref{sect.phot} to ensure that the quality of the
photometry is consistent with that of the general stellar population. The
cumulative distributions of the vertical distance from the MS are shown in
Fig.~\ref{fig.msrad} for the 2 CMDs, where we plot the distributions for
the active binaries, the AGB01 binaries and for all stars, excluding
outliers with offsets greater than $\pm$ 1.5 mag (1.1\% and 2.6\% of the
total stellar populations in the $U$ vs $U-V$ and $V$ vs $V-I$ CMDs).

In the $U$ vs $U-V$ CMD the set of active binaries includes the IDs for
W12, W14, W18, W22, W41, W47, W69, W137, W182, \& W184. The distribution of
MS offsets for these objects is consistently brighter than the general
stellar population, as expected if they contain a significant number of
binaries with reasonably bright secondaries (the median $\delta mag$
offsets for the X-ray detected active binaries, the AGB01 binaries and the
total stellar population are --0.27, -0.07 and 0.001 respectively).  Using
the KS-test there is a 99.21\% probability that the distributions for the
active binaries and the general stellar population are different (in this
and in subsequent use of the KS-test we quote the probability that the
distributions are different). This distribution is also somewhat brighter
on average than that for the AGB01 binaries (KS probability = 62.6\%),
although the faint tail of the AGB01 distribution contains several likely
spurious $U$ values (from visual examination of the images) that skew the
distribution to fainter values. In $V$ vs $V-I$ the active binary and AGB01
distributions are consistent with each other (KS probab. = 41.3 \%) and
they are both inconsistent with the general stellar distribution at $>$
99.999\% level (here the sample includes the binaries included in $U$ vs
$U-V$ plus the IDs for W26, W59, W66, \& W68). The ID for W23 is too near
the vertical part of the MS to be included in this sample, and W94\opt\ and
W121\opt\ are too crowded for reliable photometry (the light contamination
from neighboring stars is too high; see \S~\ref{sect.phot}). Here, the
median $\delta mag$ offsets for the X-ray detected active binaries, the
AGB01 binaries and the total stellar population are --0.42, -0.47 and 0.014
respectively.

The other active binaries are too bright to be included in this
analysis. The IDs for W92, W163 and W167 are blue stragglers, while W9\opt\
and W75\opt\ (BY~Dras) are found several tenths of a magnitude above the
MSTO in both CMDs (see Fig.~\ref{fig.msto}), consistent with them being
binaries. The IDs for W3 and W72 have been classified as red stragglers
(AGB01). The ID for W14 is found well to the red of the MS in both CMDs and
is also a red straggler candidate (AGB01 found $V$=17.63, $U-V$=0.87,
$V-I$=0.70; we find $V$=17.61, $U-V$=0.87, $V-I$=0.89). Note that W14\opt\
also appears to be significantly redwards of the MS in the GO-7503 CMD
(Fig.~\ref{fig.meycmds}), consistent with the red straggler explanation.
The IDs for W38 (an eclipsing binary) and W73 and W76 (both BY~Dras) are
all found close to the $U$ vs $U-V$ ridge-line but slightly to the red of
the subgiant branch in $V$ vs $V-I$, a possible hint to further red
straggler type behavior (formally W38\opt\ is found within the red
straggler box shown in Fig.~\ref{fig.msto}).  Finally, in the $V$ vs $V-I$
CMD, W69\opt\ is 1.56 mag brighter than the MS, but this object is not
obviously a red straggler.

Active binaries generally should not have UV-bright components. This is
consistent with the lack of detection (as CV candidates) of any of the 6
\cha\ sources identified with active binaries (W14, W18, W26, W41, W73,
W75) lying in the far-UV STIS image by Knigge et al. (2002).  Knigge et
al. (2002) suggest that the optical ID for W75 is 1V36 (V36 using their
nomenclature), but an astrometric analysis shows that 1V36 is 4.35-$\sigma$
away from W75 and is therefore unlikely to be the ID for W75 (our proposed
ID W75\opt\ is 2.3-$\sigma$ away).  However, the \cha\ image
(Fig.~\ref{fig.possblue-fchart}) does show enhanced counts near 1V36,
possibly because it corresponds to a weak source.  Deeper follow-up images
with \cha\ may confirm this possible detection.  Further discussion of this
interesting object will be given in Paper~II.

We have not undertaken a general search of the GO-7503 dataset for
variability (this dataset is clearly greatly inferior to the GO-8267 data
for such searches) but a few likely and possible active binaries have been
found in the GO-7503 data based on CMD position alone. By far the best
candidate is W43\opt (see Fig.~\ref{fig.meycmds}), a red straggler
previously reported as a BY~Dra by GHE01a. A possible counterpart to W64 is
found above the MS and is a good active binary candidate, while W54 is a
possible RS~CVn, consistent with the Ferraro et al. (2001a) detection of a
nearby bright object. Note that W64 is a bright source (164 counts) but
shows no evidence for variability in the \cha\ observation, unlike
W47. However, it would probably have been detected in the ROSAT
observations if it had been this bright, suggesting long-term variability.

\subsection{MSPs}
\label{sect.msps}

Of the 20 MSPs reported by Camilo et al. (2000) and Freire et al. (2001),
13 are in binary systems and have companions that are potentially
detectable in the optical, and all but 47~Tuc~J fall in at least one of the
GO-8267 or GO-7503 FoVs. The secondary star masses are predicted by Camilo
et al. (2000) to either fall in a relatively high mass range of
$\sim$0.15--0.2\mdot\ (47~Tuc~E, H, Q, T, S, U, W) or 0.02--0.03\mdot\
(47~Tuc~I, J, O, P, R). The MSP 47~Tuc~V is believed to have a relatively
massive companion, but more detections of this object are required to give
a mass estimate.

A previous search (using the GO-8267 data) was made for blue optical
counterparts to binary MSPs (Edmonds et al. 2001) in the cases where the
MSPs were detected as X-ray sources using {\tt WAVDETECT} (see GHE01a and
Grindlay et al. 2002) and the radio sources have timing positions (47~Tuc
E, H, I, O, and U). One obvious optical counterpart to an MSP in this
cluster (47~Tuc~U), with an associated \cha\ source (W11) was discovered by
Edmonds et al. (2001). This blue star is almost certainly a He~WD and shows
small amplitude orbital variations at the same period and phase as the
radio binary.  We have found no other unambiguous association of blue stars
and MSPs in searching the optical images at the positions of the
radio/X-ray MSPs.

For 47~Tuc~E (\cha\ source W7), several stars, some of them relatively
bright, are found within the 3$\sigma$ error circle. The power spectra of
two nearby ($<3\sigma$) MS stars with $V$=17.9 \& 18.5 show no evidence for
periodic or non-periodic variability. Difference images\footnote{Difference
images are generated by subtracting from each individual image the mean
over-sampled image evaluated at the appropriate dither position; see AGB01}
were calculated for these stars and then the period and phase of the radio
MSP was used to create mean difference images within $\pm$0.07 in orbital
phase of the expected maximum and minimum light. These showed no evidence
for a periodic signal (ie a stellar shaped source).  The MSP 47~Tuc~H (W74)
is found in a relatively blank region of 47~Tuc and we are able to set
upper limits in $U$ and $V$ of 25.8 and 23.7 respectively. No stars with
non-MS colors or variability were found within the error circle of the MSP
47~Tuc~O (W39).

To look for binary MSP counterparts in the cases where no X-ray source was
detected with {\tt WAVDETECT} (47~Tuc Q and T), we used the astrometry
techniques given in \S~\ref{sect.ast} (and used in Grindlay et al. 2002) to
shift the radio coordinate frame onto the X-ray coordinate frame using the
6 relatively isolated MSPs detected with \cha. The resulting positional
errors for the 16 MSPs after making these linear fits were dominated by the
relatively large errors (0\farcs1-0\farcs15) in the X-ray positions (the
radio errors are a few milliarcseconds).  Then the corrected X-ray
positions were transformed to the \hst\ coordinate frame, on a chip-by-chip
basis, using the large number of optical counterparts to X-ray sources,
with the smaller errors for this transformation (compared to the radio to
X-ray transformation) propagated into the final MSP errors.

\begin{figure}
%\vspace*{-0.5cm}
%\hspace*{-0.1cm}
%\epsfig{file=../../paper-gillil-pc1-wf2.eps,width=18.2cm}
\epsscale{1.0}
\plotone{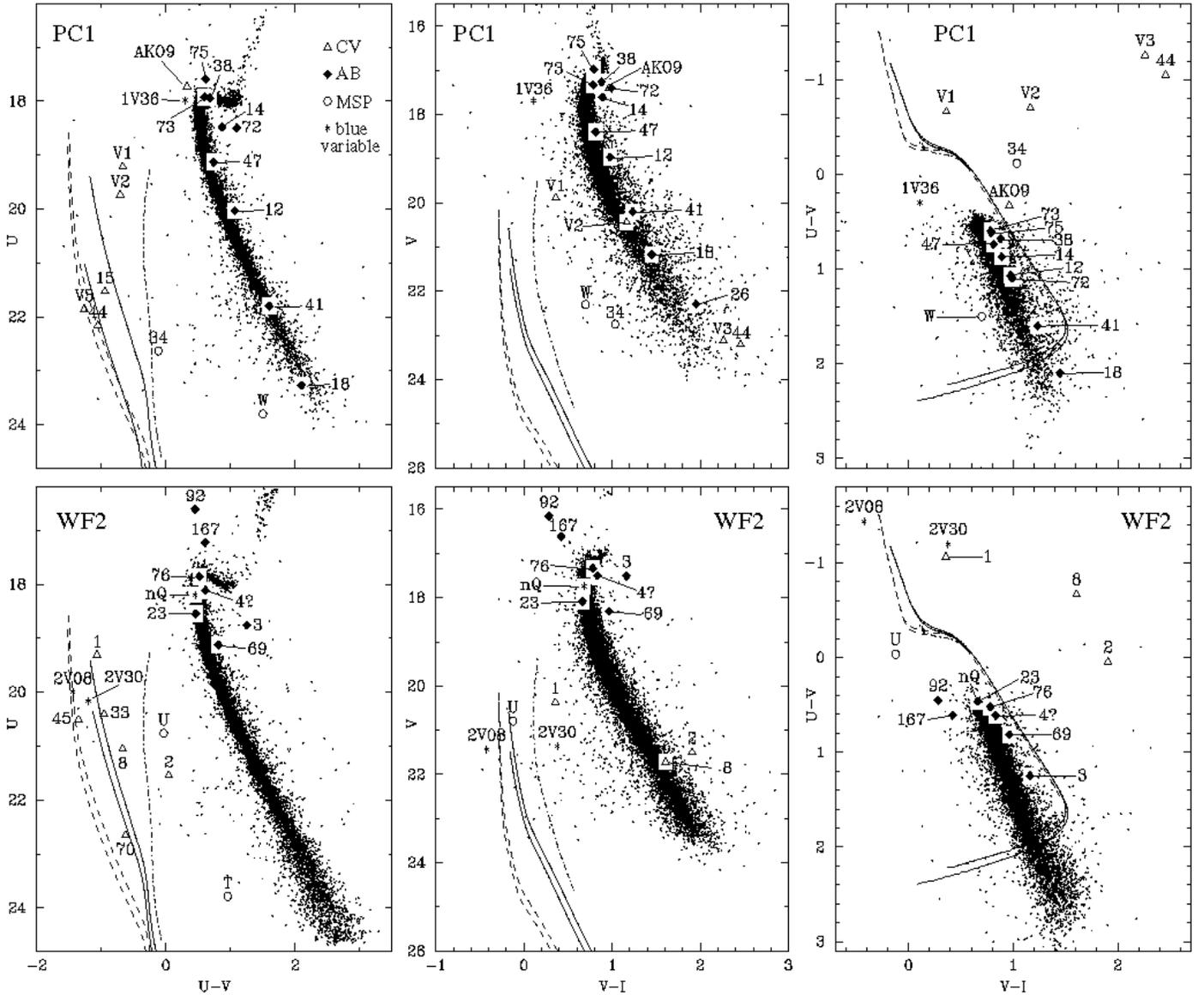}
%\vspace*{0.2cm}
\caption{$U$ vs $U-V$, $V$ vs $V-I$ and $U-V$ vs $V-I$ CMDs for the PC1 and
WF2 chips for GO-8267. Optical counterparts for \cha\ sources are shown
with numerals (W\#), except for V1 (W42), V2 (W30) and V3 (W27), the
optical counterparts to MSPs 47~Tuc~U (`U') and 47~Tuc~W (`W') and the
possible counterpart to 47~Tuc~T. The likely variable near MSP 47~Tuc~Q is
shown as `nQ', and the blue variables from AGB01 that appear to be X-ray
dim (or in the case of 1V36 a possible weak source) are also shown (see
Fig.~\ref{fig.possblue-fchart}).  The two dashed lines show 0.5 \mdot\ and
0.6 \mdot\ WD cooling tracks from Bergeron, Wesemael, \& Beauchamp (1995),
the solid lines show He~WD cooling tracks from Serenelli et al. (2001) for
0.196 \mdot\ and 0.406 \mdot\ models and the dot-dashed line the 0.169
\mdot\ model. }
%\vspace*{0.5cm}
\label{fig.gill-pc1wf2}
\end{figure}

\begin{figure}
%\vspace*{0.5cm}
%\hspace*{-0.1cm}
%\epsfig{file=../../paper-gillil-wf3-wf4.eps,width=18.2cm}
\plotone{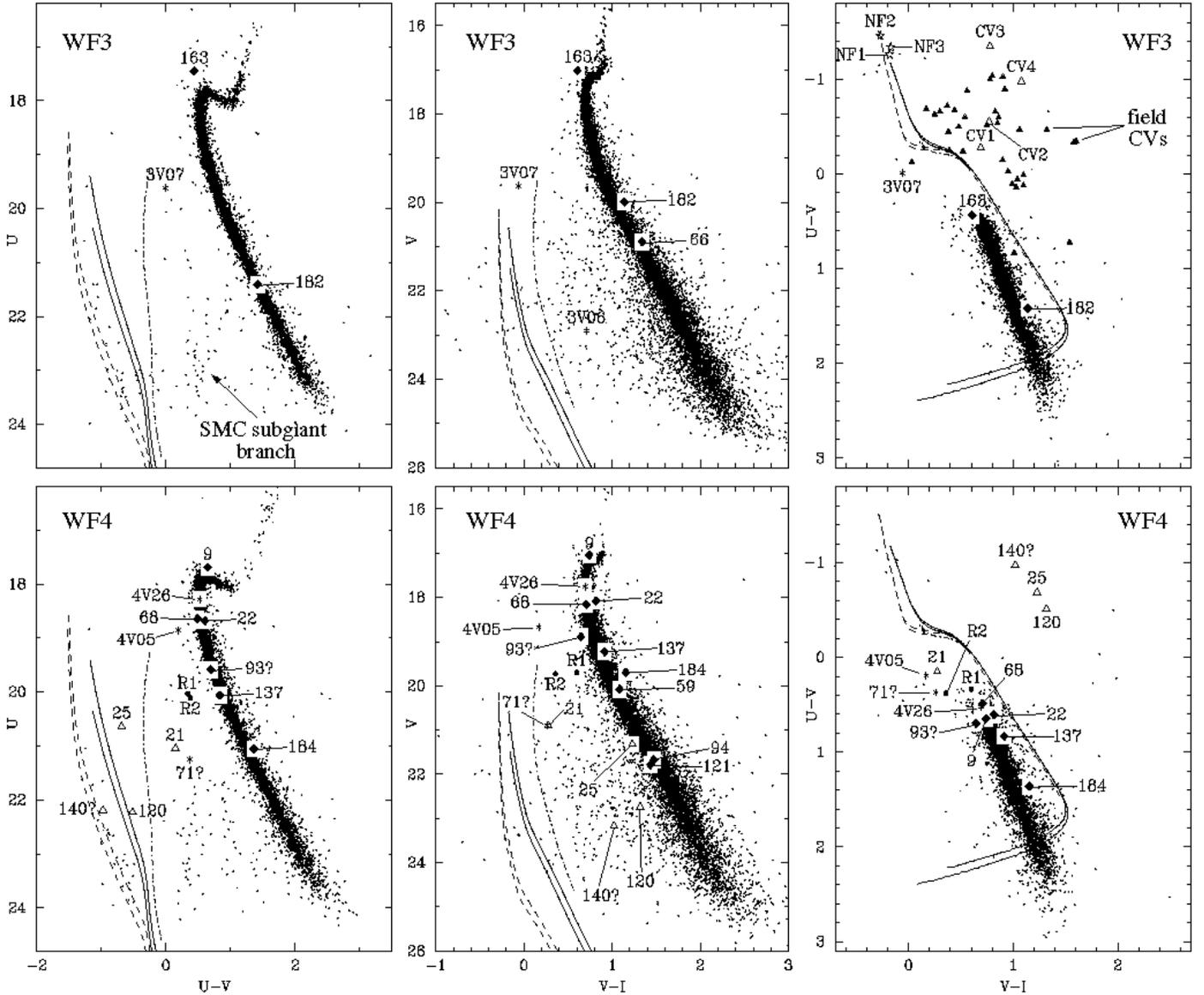}
%\vspace*{0.2cm}
\caption{$U$ vs $U-V$, $V$ vs $V-I$ and $U-V$ vs $V-I$ CMDs for the WF3 and
WF4 chips for GO-8267. For WF3 we plot colors for the 4 brightest CVs in
NGC~6397 (CV1-CV4) and the 3 brightest He~WDs in NGC~6397 (NF1-NF3, where
`NF' denotes non-flickerer) from Cool et al. (1998).  We also plot (with
filled-in triangles) the field CVs with $UVI$ colors from Zwitter \& Munari
(1995), Zwitter \& Munari (1996), and Munari \& Zwitter (1998), plus two RR
Lyraes (labeled with filled circles) from the SMC (unpublished work from
the authors). Note: (1) the relative lack of CVs falling below the WD
cooling tracks in the color-color plot, (2) the populations of cluster WDs
detectable in the $U$ vs $U-V$ CMDs at $U-V<0$ and (3) the subgiant branch
of the SMC visible in the $U$ vs $U-V$ CMDs (as labeled).}
%\vspace*{0.5cm}
\label{fig.gill-wf3wf4}
\end{figure}

\begin{figure}
%\vspace*{0.5cm}
%\hspace*{-0.2cm}
%\epsfig{file=../../paper-msto.eps,width=8.5cm}
\epsscale{0.85}
\plotone{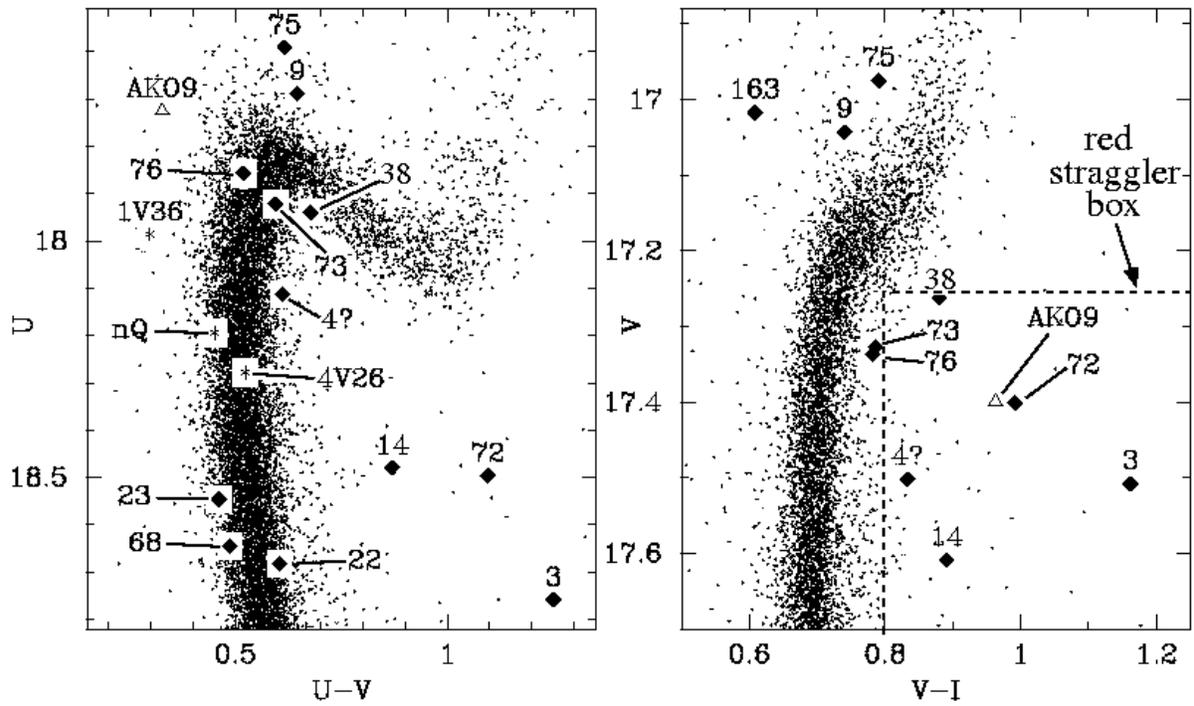}
%\vspace*{0.2cm}
\caption{A close-up of the region near the MSTO in the $U$ vs $U-V$ and $V$
vs $V-I$ CMDs for the GO-8267 data. The points for all 4 chips are plotted
together, and a portion of the red straggler box from AGB01 is shown (0.8
$< V-I <$1.5; 17.25$< V <$17.75). The red stragglers from AGB01 are plotted
(W3\opt, AKO~9, and W72\opt) plus other possible red stragglers (W4\opt,
W14\opt, and W38\opt).}
%\vspace*{0.5cm}
\label{fig.msto}
\end{figure}

\begin{figure}
%\vspace*{-0.5cm}
%\hspace*{-0.1cm}
%\epsfig{file=../paper-meycmd.eps,width=15.2cm}
\epsscale{0.9}
\plotone{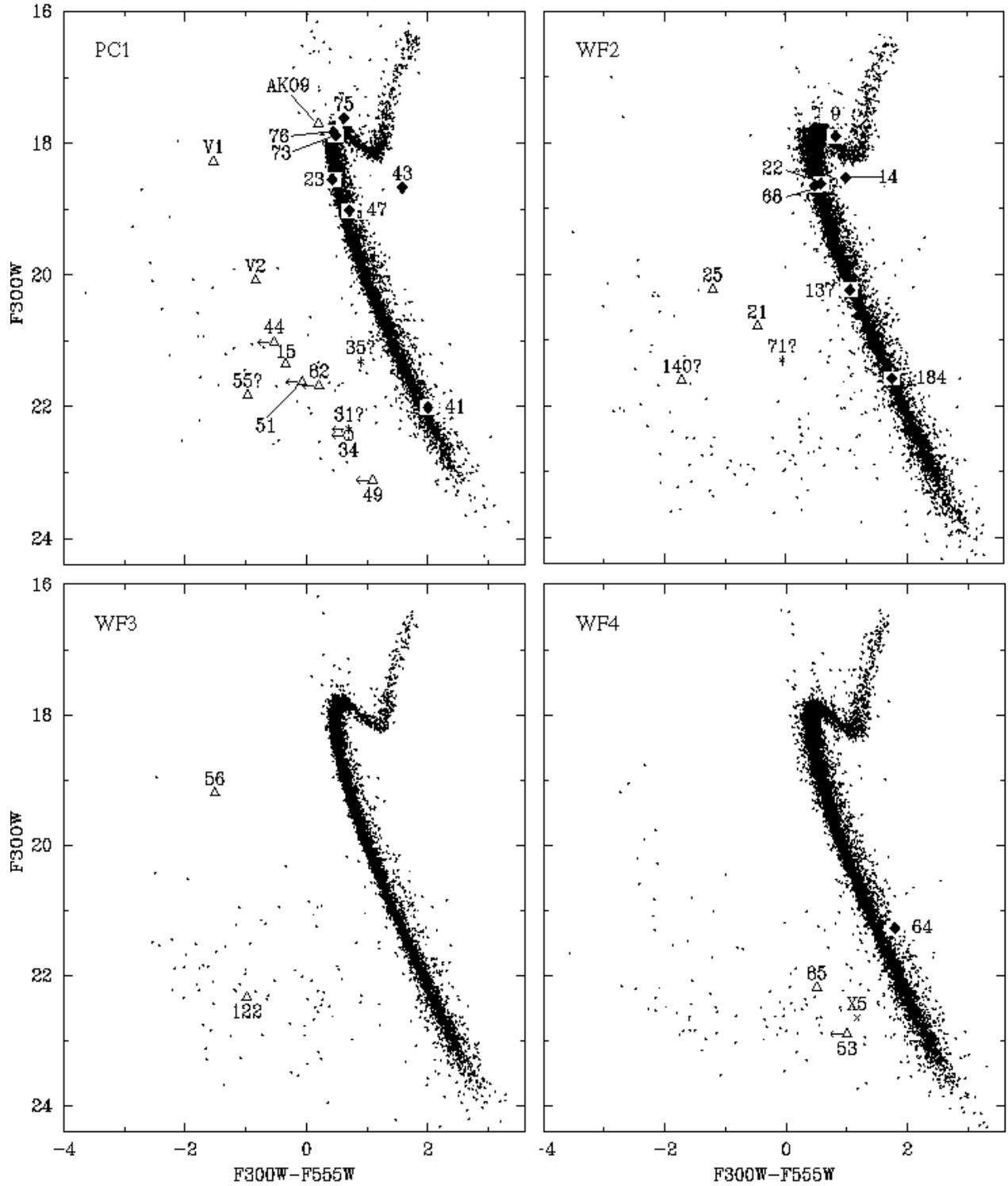}
%\vspace*{0.2cm}
\caption{F300W vs F300W-F555W CMDs for the GO-7503 dataset. Likely or
marginal counterparts to X-ray sources are labeled with numerals and
symbols (as used in Figures \ref{fig.gill-pc1wf2} and
\ref{fig.gill-wf3wf4}) in the cases where the stars are detected in both
F300W and F555W. X-ray IDs detected in F300W but not F555W are labeled at
the red limit with arrows.}
%\vspace*{0.5cm}
\label{fig.meycmds}
\end{figure}

\begin{figure}
%\vspace*{0.5cm}
%\hspace*{-0.1cm}
%\epsfig{file=../../paper-xfig-close-colcol.eps,width=8.0cm}
\epsscale{0.75}
\plotone{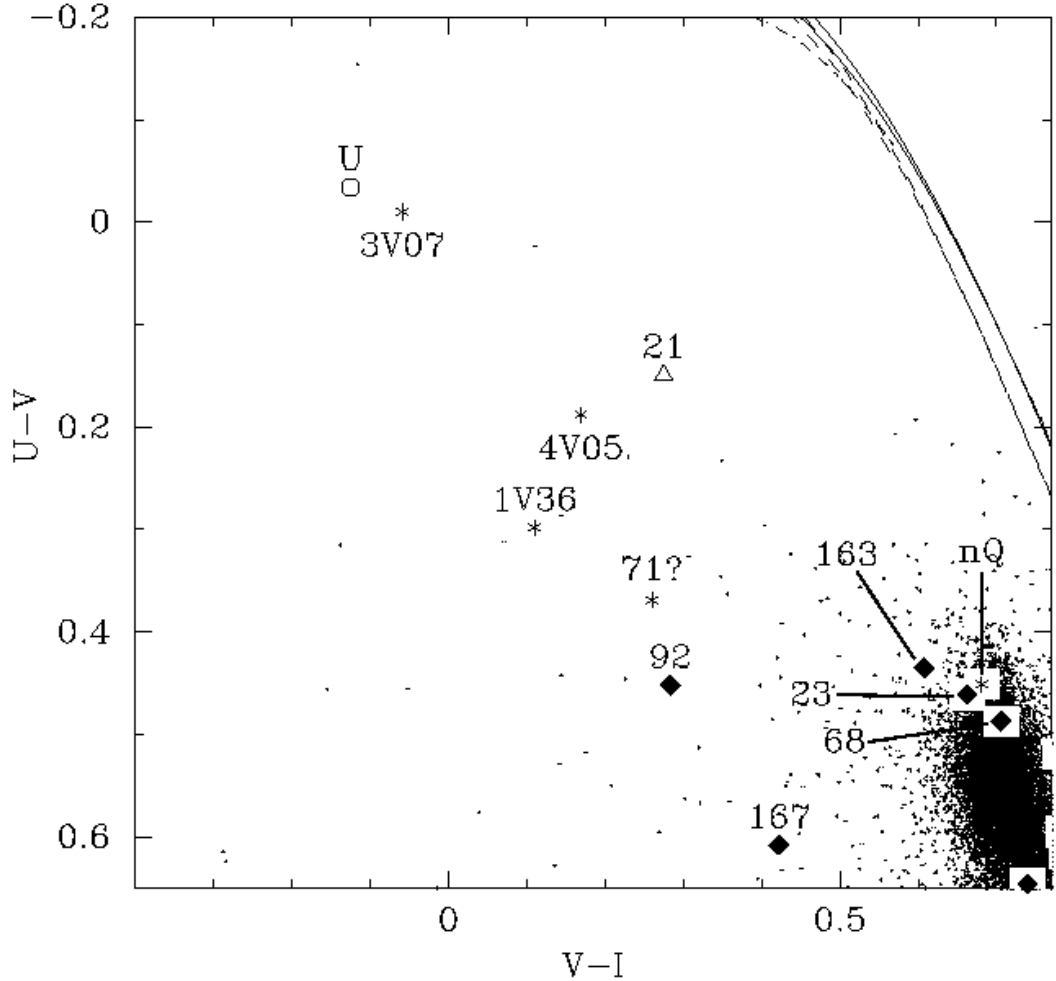}
%\vspace*{0.2cm}
\caption{A close-up of the $U-V$ vs $V-I$ CMD for all 4 WFPC2 chips for
GO-8267, showing 4 of the blue variables (1V36, 3V07, 4V05, and the blue
variable near W71) that are not obviously associated with {\tt WAVDETECT}
X-ray sources, plus W21\opt\ and the counterpart to 47~Tuc~U. The stars
W92\opt, W163\opt\ and W167\opt\ are blue stragglers.}
%\vspace*{0.5cm}
\label{fig.close-colcol}
\end{figure}

\begin{figure}
%\vspace*{0.5cm}
%\hspace*{-0.2cm}
%\epsfig{file=/data/edmonds/chandra/47tuc/paper-ms-rad.eps,width=8.5cm}
\epsscale{1.0}
\plotone{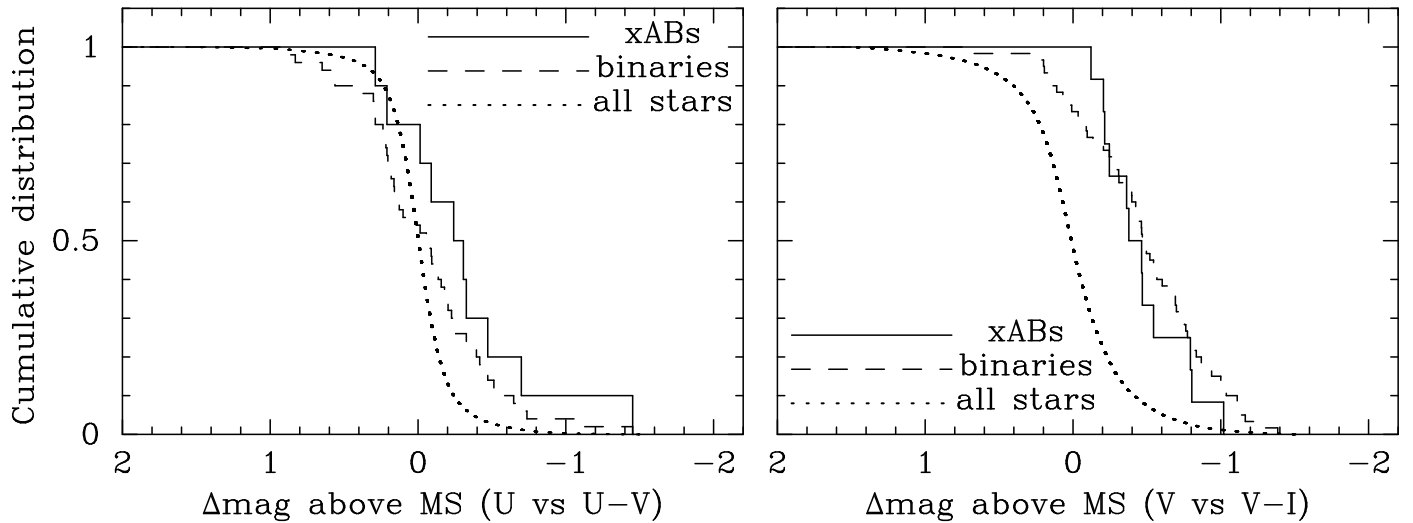}
%\vspace*{0.2cm}
\caption{Offsets above the main sequence in the $U$ vs $U-V$ ($U>18.4$) and
the $V$ vs $V-I$ ($V>18$) CMDs for the AGB01 binaries (`binaries'), the
X-ray detected active binaries (`xABs') and the overall population of stars
(`all stars').}
%\vspace*{0.5cm}
\label{fig.msrad}
\end{figure}

\begin{figure}
%\vspace*{0.5cm}
%\hspace*{-0.2cm}
%\epsfig{file=../fchart/paper-47tuct-fchart.eps,width=8.5cm}
\epsscale{0.8}
\plotone{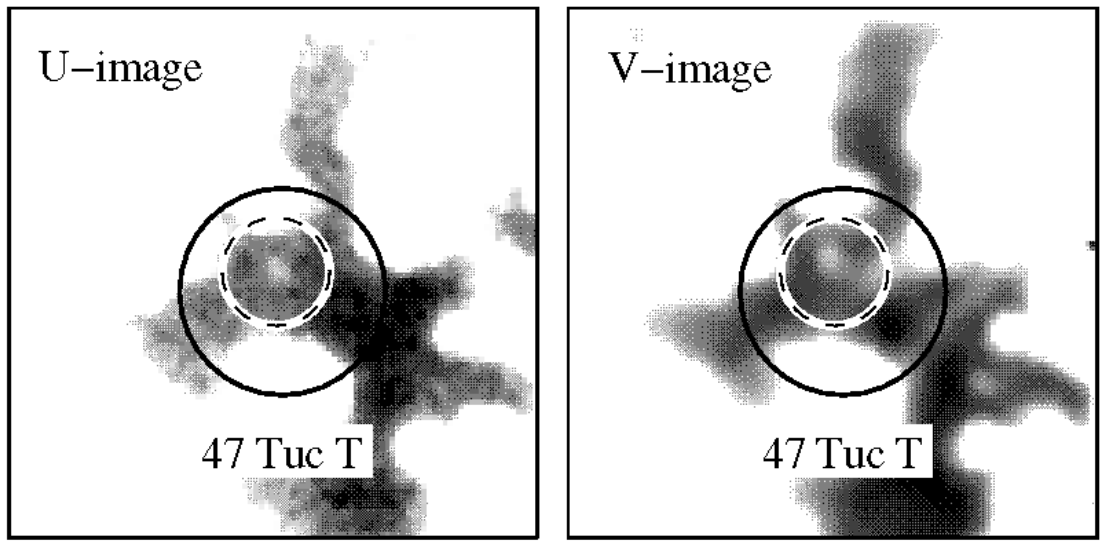}
%\vspace*{0.2cm}
\caption{Finding charts in $U$ and $V$ from the GO-8267 dataset for the
possible optical counterpart to the MSP 47~Tuc~T. The solid line shows the
4$\sigma$ error circle (1$\sigma$=0\farcs11) for the MSP and the dotted
line shows the faint star detected in the $U$-band image. The faint object
detected in the $V$-band image is clearly a different star because of the
significant spatial shift.}
%\vspace*{0.5cm}
\label{fig.47tuct}
\end{figure}

Using this astrometry, one likely, very faint ($U\sim 23.8$; see
Fig.~\ref{fig.gill-pc1wf2}) blue star is found on the WF2 chip only
0.11$''$ (0.9$\sigma$) from the position of 47~Tuc~T (Fig.~\ref{fig.47tuct}
shows a finding chart). No corresponding star is visible in the $V$ or $I$
images in the PSF wings of a neighboring red star, showing that the
possible MSP counterpart is indeed blue (faint MS stars will appear to be
much fainter in the $U$ image than in the $V$ or $I$ image). If this star
is a He~WD then using the He~WD cooling curves of Serenelli et al. (2001)
corrected to the 47~Tuc distance and reddening (see
Fig.~\ref{fig.gill-pc1wf2}) we infer that the star has a cooling age
$\lesssim 0.5$ Gyr if the WD mass is $\gtrsim$0.2 \mdot, and a cooling age
of $\sim$10 Gyr if the WD mass is $\sim0.17$\mdot\ (the lowest mass model
of Serenelli et al 2001). Since perhaps only future generation telescopes
are likely to produce deeper $V$ images of 47~Tuc than this dataset, it may
be some time before useful constraints on the WD cooling age are made from
optical color information. A possible blue star also exists within 0.09$''$
(0.82$\sigma$) of 47~Tuc~I (W19), however in this case the possible
$U$-bright star is much more crowded than the candidate near 47~Tuc~T, and
is only a marginal candidate at best. The PSF is not unambiguously stellar
and could, e.g. be the combination of light from several unresolved stars.
The derived $U$ magnitude is $\sim$24 and the extreme crowding is likely to
limit useful follow-up observations at optical wavelengths, but UV
observations with ACS/HRC may provide useful constraints. The detection of
an optical counterpart to 47~Tuc~I would be of great interest, since this
MSP has a very low mass companion (no optical companions for such objects
have been detected in globular clusters).

For 47~Tuc~Q there is one bright ($V=17.7$) star only 0\farcs24
(1.3$\sigma$) away from the corrected MSP position, and this star is found
on the MS near the turnoff (`nQ' in Fig.~\ref{fig.gill-pc1wf2}). This shows
no evidence for a power spectrum signal at the 1.189 day binary period
detected in the radio, but it does show evidence for a signal in the
$V$-band with period$ = (0.391359 \pm 0.001700)$ days and amplitude =
$0.00144 \pm 0.00017$ (8.5$\sigma$). This period is $\sim3\sigma$ away from
being one third of the period of 47~Tuc~Q. No interesting signal is seen in
$I$.  Since the binary orbit of 47~Tuc~Q is very close to circular (Freire
et al. 2001), this shorter period is unlikely to represent evidence for
enforced, rapid rotation at periastron in a highly elliptical
orbit. Therefore, we believe this possible optical counterpart is most
likely a chance coincidence.

The binary MSP 47~Tuc~S (W77), thought to have a He~WD companion (Camilo et
al. 2000), falls in the GO-7503 FoV but no blue stars are visible by
comparing the F300W and F555W images.  Four of the MSPs (47~Tuc~P, R, V and
W) from the sample of Freire et al. (2001) do not have timing positions,
and one of these (47~Tuc~W) has been identified with the optically variable
star \wopt\ by Edmonds et al. (2002b). The prospects of further discoveries
like this are discussed in Paper~II.

\subsection{Unidentified X-ray sources}
\label{sect.unid}

There are only a small number (8) of moderately bright ($>20$ ct) \cha\
sources in the GO-8267 source list without plausible optical counterparts.
The source W46 is almost certainly a qLMXB based on the X-ray luminosity
and spectrum (GHE01a and Heinke et al. 2003) but has no plausible optical
counterpart (Edmonds et al. 2002a). The possible CV ID for W35 was
discussed in \S~\ref{sect.cvphot}. The source W16 (102 cts) falls in a very
crowded part of the \hst\ image (it lies 0.6$''$ from a bright, variable
red giant).  Its nearest counterparts in the $U$ images are at 1.2, 3.8 and
3.9 $\sigma$ with $U$ mags of 21.4, 18.3 and 20.5 respectively, but the
effects of the giant star severely limit results in $V$ and $I$. Because of
the hard X-ray spectrum and its reasonably high luminosity this source is
most likely a CV. Finally, W17 is discussed in Paper~II as a possible CV or
qLMXB.

The sources W20, W24, W32 and W37 all have bright, but apparently
non-variable stars within 3$\sigma$ of the X-ray positions (W20 has a
$V$=17.6 star at 0.44$\sigma$; W24 a $V$=16.0 star at 1.35$\sigma$; W32
a $V$=17.5 star at 1.48$\sigma$; W37 was discussed in \S~\ref{sect.abs}).
Presumably these bright stars could be active binaries with low
inclinations giving a small observed amplitude for orbital variations (and
the star near W24 is saturated, so the time series have lower quality than
available for MSTO stars), or they could be overpowering the light of
fainter optical IDs, such as CVs.  Also, the bright star near W24 is found
on the blue side of the giant branch (based on preliminary photometry of
short exposures), suggesting binarity.  Although none of these objects are
suggested as CV candidates by Knigge et al. (2002), their CMD uses a color
based on far-UV and F336W magnitudes, and the light contamination from the
nearby MS stars and subgiants may prevent the detection of fainter objects
in F336W that represent the true counterparts. However, MSTO stars and
subgiants are much fainter in the far-UV and therefore, close examination
of the images of Knigge et al. (2002), and spectroscopic and variability
studies of nearby UV stars should be useful in searching for CV
counterparts in the UV.  These 4 \cha\ sources may also have MSP
counterparts, but they are all brighter than the most luminous MSP and the
faintest (W37) is a variable X-ray source.

Several of the fainter ($<20$ ct), unidentified sources, have saturated
stars nearby that may have prevented the detection of faint objects like
CVs. The source W10 has a saturated star ($V$=15.3) within 0\farcs27
(2.4$\sigma$), W31 has a saturated star ($V$=15.9) within 0\farcs23
(2.1$\sigma$), W98 has 3 saturated stars ($V$=15.1-16.0) within 0\farcs4
(2.0$\sigma$), W115 has a saturated star ($V$=17.0) within 0\farcs3
(2.1$\sigma$), W141 has a saturated star ($V$=14.0) within 0\farcs22
(1.5$\sigma$), and W168 has a saturated star ($V$=13.6) within 0\farcs16
(0.7$\sigma$). Several of these examples (especially W168 and W141) may
represent the detection of the saturated star itself, most likely as an RS
CVn.

\section{Discussion}
\label{sect.disc}

\subsection{Cataclysmic variables}
\label{sect.cvdisc}

We have discovered (or confirmed) optical counterparts for 22 \cha\ sources
that are likely CVs as summarized in Tables~\ref{tab.8267} and
\ref{tab.7503} (we have included V3 in this list, but it may be a qLMXB, as
discussed in Paper~II). Edmonds et al. (2002b) tentatively classified
W34\opt\ as an MSP companion based on the similarity of its $V$ and $I$
light curves with the 47~Tuc~W companion, but it may also be a CV.
Alternatively, a small number of MSPs may be included in our CV sample, as
discussed in Paper~II. Among the marginal candidates given in
Tables~\ref{tab.8267} and \ref{tab.7503} we believe that W140\opt\ and
W55\opt\ (see Fig.~\ref{fig.marg-fchart}) have the greatest chances of
being CVs, although the chance of astrometric coincidences is now
significant (4.1\% and 10.8\%).

This is easily the largest sample of CVs detected in a single globular
cluster. The photometric properties of these stars are summarized as: (1)
they all have $U-V$ or F300W-F555W colors that place them well bluewards of
the MS, with the exception of AKO~9 (with its bright secondary); (2) the
optical IDs for W2, W8, W25, V2, V3, W44 and W120 all lie on or close to
the MS in $V$ vs $V-I$ (similar behavior is seen in the CVs discovered in
NGC~6397 and NGC~6752), while only V1, W1\opt\ and W21\opt\ and possibly
W34\opt\ are found well to the blue of the MS. These results show that the
secondaries generally dominate the optical light and that the accretion
disks are relatively faint, suggesting that these systems have low
accretion rates.  Strong support for this hypothesis is given in Paper~II,
where the periods for some of these objects, and their absolute magnitude,
X-ray luminosity and \fxfopt\ distributions will be presented.

Unfortunately, because of crowding in the \hst\ images, our sample of CVs
does not reach the very faint absolute magnitudes (\mv $>$10) discussed by
Townsley \& Bildsten (2002) in their modeling of low accretion rate CVs
with periods below the period gap. For making direct comparison with the
$V$ and $I$ magnitudes predicted by Townsley \& Bildsten (2002), matters
are even worse because: (1) some of the faintest CVs in the GO-8267 FoV
lack $I$ magnitudes (e.g. W15\opt\ \& W70\opt), and (2) the CVs found only
in the GO-7503 FoV all lack $I$ magnitudes (and, in many cases, F555W
magnitudes). A couple of the faintest CV candidates (W34\opt\ \& W140\opt)
do have bluer colors than the MS, suggesting a greater contribution from
the WD, as predicted by Townsley \& Bildsten (2002) at faint
magnitudes. However, W34\opt\ could be an MSP, and W140\opt\ is only a
marginal CV candidate, and even if they are CVs, their bluer colors could
be explained by either relatively bright accretion disks or lower
inclinations than average.

We do not find convincing evidence for X-ray emission from the blue
variables discussed by AGB01, although one of these objects (1V36) may be a
weak, crowded X-ray source; see also the discussions of the UV and optical
properties of 1V36 in Ferraro et al. (2001a) and Knigge et
al. (2002). Possible explanations for these blue variables will be
discussed in Paper~II.

\subsection{Active binaries}
\label{sect.abdisc}

Adding up the total number of active binary candidates in
Tables~\ref{tab.8267} and \ref{tab.7503} (counting only W43 and W64 of the
GO-7503 data candidates) we have observed a total of 29 likely active
binaries. Except for W43\opt\ and W64\opt, all of these are found in the
GO-8267 FoV and show statistically significant, mostly periodic,
variability (see Paper~II). They are mostly found on, or slightly above,
the MS or subgiant ridge-line except for W69\opt\ and a handful of red
stragglers or red straggler candidates.  Several marginal active binaries
have also been found. The total sample of active binaries in 47~Tuc above
our X-ray detection threshold will inevitably be larger, and in particular
a number of MS-MS binaries that were outside the GO-8267 FoV should be
found in the GO-7503 FoV.

\subsection{Summary}
\label{sect.summ-prosp}

Almost certain matches are available for 45 of 78 X-ray sources within the
extensive \hst/WFPC2 imaging of program GO-8267 on the basis of tight
astrometric pairing with optical sources showing unusual CMD locations
and/or clear optical variability.  Six of the remaining sources in the
GO-8267 field are solitary MSPs or binary MSPs (with likely He~WD
companions) from Camilo et al. (2000), and one of the remaining sources is
a qLMXB (GHE01a; Edmonds et al. 2002a; Heinke et al. 2003), and each of
these neutron star systems are expected to have faint or extremely faint
optical counterparts.  Of the remaining 26 sources, $\sim$15 are
problematic due to astrometric coincidence with regions of low data quality
(e.g. covered by saturation trails or extremely crowded; see Paper~II for
more details), and $\lesssim$4 lack optical IDs in clean regions of optical
coverage (here the working explanation is that these sources are MSPs and
the optical counterparts fall below the detection limit of even the
extensive \hst/WFPC2 observations available to us).  Of the 84 X-ray
sources found within the GO-7503 FoV, 52 are also found in the GO-8267 FoV,
and of the remaining 32 sources, 10 have optical identifications.

Paper~II will present the time series results for these optical
identifications, plus a detailed analysis section, including a study of the
spatial distribution of the optical identifications, and the distributions
of their absolute magnitudes, X-ray luminosities and X-ray to optical flux
ratios. This will be followed by an overall interpretation of the results.

\acknowledgments

This work was supported in part by STScI grants GO-8267.01-97A (PDE and
RLG) and HST-AR-09199.01-A (PDE). We acknowledge several helpful comments
from the referee, Christian Knigge.

\clearpage

%%--------------------------FIGURE----------------------------------
%%-------------------------------------------------------------------

%\clearpage 

%%--------------------------FIGURE----------------------------------
%%-------------------------------------------------------------------

%\clearpage 

%%--------------------------FIGURE----------------------------------
%%-------------------------------------------------------------------

%\clearpage 

%%--------------------------FIGURE----------------------------------
%%-------------------------------------------------------------------

%\clearpage 

%%--------------------------FIGURE----------------------------------
%%-------------------------------------------------------------------

%\clearpage

%%--------------------------FIGURE----------------------------------
%%-------------------------------------------------------------------

%\clearpage 

%%--------------------------FIGURE----------------------------------
%%-------------------------------------------------------------------

%\clearpage 

%%--------------------------FIGURE----------------------------------
%%-------------------------------------------------------------------

%\clearpage

%%--------------------------FIGURE----------------------------------
%%-------------------------------------------------------------------

%\clearpage 

%%--------------------------FIGURE----------------------------------
%%-------------------------------------------------------------------

%\clearpage 

%%--------------------------FIGURE----------------------------------
%%-------------------------------------------------------------------

%\clearpage

%%--------------------------FIGURE-----------------------------------
%%-------------------------------------------------------------------

% ---- Table 1-------------------------------------------------------------------

\tabletypesize{\scriptsize}

\begin{deluxetable}{cllrrr}
\tablecolumns{6}
\tablewidth{0pc}
\tablecaption{\cha\ sources outside GHE01a list but within \hst\ FoV}
\tablehead{\colhead{\cha\ source} & \colhead{RA} & \colhead{Dec} & 
  \colhead{medcts\tablenotemark{a}} & \colhead{sfcts\tablenotemark{b}} & 
  \colhead{hdcts\tablenotemark{c}} \\
\colhead{number (W\#)} & \colhead{} & \colhead{} & \colhead{} & \colhead{} & \colhead{} } 
\startdata
 115 & 00:24:17.24(2) & $-$72:04:46.3(1)  &   3.6  &   3.8  &   0.0  \\ 
 120 & 00:24:11.12(1) & $-$72:06:20.13(5) &  66.2  &  38.6  &  33.5  \\ 
 121 & 00:24:05.16(3) & $-$72:06:04.3(1)  &  12.6  &   8.8  &   0.0  \\ 
 122 & 00:24:03.86(1) & $-$72:06:21.80(7) &  46.2  &  27.7  &  20.5  \\ 
 137 & 00:24:18.68(4) & $-$72:06:02.8(1)  &   3.8  &   0.0  &   0.0  \\ 
 140 & 00:24:09.22(2) & $-$72:05:43.7(1)  &   6.7  &   4.8  &   0.0  \\ 
 141 & 00:24:07.19(3) & $-$72:04:46.33(8) &   5.0  &   5.2  &   0.0  \\ 
 142 & 00:24:04.27(3) & $-$72:04:20.2(1)  &   8.3  &   0.0  &   0.0  \\ 
 145 & 00:23:54.60(3) & $-$72:06:07.7(0)  &   2.9  &   0.0  &   0.0  \\ 
 163 & 00:24:25.21(4) & $-$72:05:07.4(1)  &   6.8  &   6.8  &   0.0  \\ 
 167 & 00:24:10.24(3) & $-$72:04:01.7(2)  &   5.7  &   0.0  &   0.0  \\ 
 168 & 00:24:05.96(5) & $-$72:03:46.5(0)  &   2.9  &   0.0  &   0.0  \\ 
 169 & 00:24:03.73(0) & $-$72:05:17.0(0)  &   1.6  &   0.0  &   0.0  \\ 
 182 & 00:24:19.43(3) & $-$72:04:31.0(1)  &   2.7  &   2.8  &   0.0  \\ 
 184 & 00:24:10.52(4) & $-$72:05:28.8(1)  &   4.6  &   0.0  &   0.0  \\ 

\tablenotetext{a}{medcnts = counts in 0.5--4.5 keV band}
\tablenotetext{b}{sfcts = counts in 0.5--1.5 keV band}
\tablenotetext{c}{hdcnts = counts in 1.5--6.0 keV band}

\enddata
\label{tab.extra-sources}
\end{deluxetable}

% ---- Table 2-------------------------------------------------------------------

\tabletypesize{\scriptsize}

\begin{deluxetable}{ccccc}
\tablecolumns{5}
\tablewidth{0pc}
\tablecaption{Optical location and other names for \cha\ sources}
\tablehead{\colhead{\cha} & \colhead{GO-8267} & \colhead{GO-7503} & 
  \colhead{AGB01 name} & \colhead{Other names}  \\
\colhead{W\#} & \colhead{} & \colhead{} & \colhead{} & \colhead{} 
  } 
\startdata
  1 & WF2 & \nodata & WF2-V09  &   \\
  2 & WF2 & \nodata &          &   X13\tablenotemark{c} \\
  3 & WF2 & \nodata & WF2-V31  &  \\
  4 & WF2 & \nodata &          &  \\
  5 & WF2 & \nodata &          &  \\
  6 & WF4 & WF2     &          &  \\
  7 & WF4 & WF2     &          &    47~Tuc~E\tablenotemark{j} \\
  8 & WF2 & \nodata &          &  \\
  9 & WF4 & WF2     & WF4-V06  &  \\
 10 & WF2 & \nodata &          &  \\
 11 & WF2 & \nodata &          &  U\opt\tablenotemark{j},
  47~Tuc~U\tablenotemark{e} \\
 12 & PC1 & WF2     & PC1-V17  &  \\
 13 & WF2 & \nodata &          &  47~Tuc~N\tablenotemark{j} \\
 14 & PC1 & \nodata & PC1-V41  &   \\
 15 & PC1 & PC1     &          &   \\
 16 & WF2 & \nodata &          &    \\
 17 & WF2 & \nodata &          &   \\
 18 & PC1 & WF2     & PC1-V29  &   \\
 19 & WF2 & PC1     &          &    47~Tuc~G,I\tablenotemark{j} \\
 20 & PC1 & PC1     &          &   \\
 21 & WF4 & WF2     &          &   \\
 22 & WF4 & WF2     &          &   \\
 23 & WF2 & PC1     &          &   \\
 24 & PC1 & PC1     &          &   \\
 25 & WF4 & WF2     &          &    X11\tablenotemark{c} \\
 26 & PC1 & WF2     &          &    \\
 27 & PC1 & PC1     &     & V3\tablenotemark{a}, star
  \#716\tablenotemark{b}, X10\tablenotemark{c} \\
 28 & PC1 & PC1     &          &   \\
 29 & PC1 & PC1     &          & W29\opt\tablenotemark{d},
  47~Tuc~W\tablenotemark{e}  \\
 30 & PC1 & PC1 & PC1-V53 & V2\tablenotemark{f}, star
  \#5400\tablenotemark{b}, X19\tablenotemark{c} \\
 31 & PC1 & PC1     &          &    \\
 32 & PC1 & PC1     &          &   \\
 33 & WF2 & \nodata &          &   \\
 34 & PC1 & PC1     &          &   \\
 35 & PC1 & PC1     &          &   \\
 36 & PC1 & PC1 &PC1-V11 & AKO~9\tablenotemark{g},
  var.\#11\tablenotemark{h}, BSS-26\tablenotemark{b} \\
 37 & PC1 & PC1    &          &   \\
 38 & PC1 & PC1    &          &   \\
 39 & PC1 & PC1    &          &    47~Tuc~O\tablenotemark{j} \\
 40 & \nodata & PC1 &         &    \\
 41 & PC1 & PC1 & PC1-V19  &   \\
 42 & PC1 & PC1 & PC1-V47  & V1\tablenotemark{i}, UVE-3\tablenotemark{b}, 
  X9\tablenotemark{c}  \\
 43 & \nodata & PC1 &          &   \\
 44 & PC1 & PC1 &          &   \\
 45 & WF2 & \nodata &          &   \\
 46 & PC1 & PC1     &          &   \\
 47 & PC1 & PC1 & PC1-V08  & var. \#8\tablenotemark{h}   \\
 49 & \nodata & PC1 &          &   \\
 51 & \nodata & PC1 &          &   \\
 52 & \nodata & WF3 &          &   \\
 53 & \nodata & WF4 &          &   \\
 54 & \nodata & PC1 &          &   \\
 55 & \nodata & PC1 &          &   \\
 56 & \nodata & WF3 &          &    X6\tablenotemark{c}  \\
 57 & \nodata & PC1 &          &    \\
 58 & \nodata & WF4 &          & X5\opt\tablenotemark{k},
  X5\tablenotemark{c}  \\
 59 & WF4     & WF3 &          &    \\
 60 & \nodata & WF4 &          &    \\
 61 & \nodata & WF4 &          &    \\
 62 & \nodata & WF4 &          &    \\
 64 & \nodata & WF4 &          &    \\
 65 & \nodata & WF4 &          &    \\
 66 & WF3     & WF2 & WF3-V08  &    \\
 67 & WF2     & \nodata &          &    47~Tuc~D\tablenotemark{j} \\
 68 & WF4     & WF2     &          &    \\
 69 & WF2     & \nodata & WF2-V18  &    \\
 70 & WF2     & \nodata &          &    \\
 71 & WF4     & WF2     &          &    \\
 72 & PC1     & \nodata & PC1-V48  &    \\
 73 & PC1     & PC1     & PC1-V24  &    \\
 74 & WF2     & \nodata &          &    47~Tuc~H\tablenotemark{j} \\
 75 & PC1     & PC1     & PC1-V23  &   \\
 76 & WF2     & PC1     & WF2-V17  &    \\
 77 & \nodata & PC1     &          &    47~Tuc~F,S\tablenotemark{j}  \\
 78 & \nodata & WF3     &          &    \\
 79 & \nodata & PC1     &          &    \\
 80 & \nodata & PC1     &          &    \\
 81 & \nodata & WF4     &          &    \\
 82 & \nodata & PC1     &          &    \\
 84 & \nodata & WF3     &          &    \\
 85 & \nodata & WF4     &          &    \\
 86 & \nodata & WF4     &          &    \\
 87 & \nodata & WF4     &          &    \\
 90 & \nodata & WF3     &          &    \\
 91 & WF2     & \nodata &          &    \\
 92 & WF2     & \nodata & WF2-V03  &    \\
 93 & WF4     & WF2     &          &    \\
 94 & WF4     & WF2     &          &    \\
 95 & WF2     & \nodata &          &    \\
 96 & PC1     & WF2     &          &    \\
 97 & PC1     & WF2     &          &    \\
 98 & PC1     & PC1     &          &    \\
 99 & WF4     & WF3     &          &    \\
101 & \nodata & WF3     &          &    \\
102 & \nodata & WF3     &          &    \\
103 & \nodata & WF4     &          &    \\
115 & WF3     & \nodata &          &    \\
120 & WF4     & WF2     &          &    \\
121 & WF4     & WF3     &          &    \\
122 & \nodata & WF3     &          &    \\
137 & WF4     & WF2     & WF4-V03  &    \\
140 & WF4     & WF2     &          &    \\
141 & PC1     & PC1     &          &    \\
142 & WF2     & \nodata &          &    \\
145 & \nodata & WF3     &          &    \\
163 & WF3     & \nodata & WF3-V05  &    \\
167 & WF2     & \nodata &          &    \\
168 & WF2     & \nodata &          &    \\
169 & \nodata & WF2     &          &    \\
182 & WF3     & \nodata & WF3-V01  &    \\
184 & WF4     & WF2     &          &    \\

\tablenotetext{a}{Shara et al. (1996)}
\tablenotetext{b}{Ferraro et al. (2001a)}
\tablenotetext{c}{Verbunt \& Hasinger (1998)}
\tablenotetext{d}{Edmonds et al. (2002b)}
\tablenotetext{e}{Camilo et al. (2000)}
\tablenotetext{f}{Paresce \& de Marchi (1994)}
\tablenotetext{g}{Auriere et al. (1989)}
\tablenotetext{h}{Edmonds et al. (1996)}
\tablenotetext{i}{Paresce et al. (1992)}
\tablenotetext{j}{Edmonds et al. (2001)}
\tablenotetext{k}{Edmonds et al. (2002a)}

\enddata
\label{tab.other-names}
\end{deluxetable}

% ---- Table 3-------------------------------------------------------------------

\tabletypesize{\scriptsize}

\begin{deluxetable}{rcrrcccrrccll}
\tablecolumns{13}
\tablewidth{0pc}
\tablecaption{\hst\ (GO-8267) data for counterparts to 47 Tuc \cha\ sources}
\tablehead{\colhead{Source} & \colhead{Chip \#} &
 \colhead{$\Delta$RA} & \colhead{$\Delta$Dec} & \colhead{signif.} &  \colhead{chance} &
 \colhead{$V$} & \colhead{$U-V$} & \colhead{$V-I$} & \colhead{RA\tablenotemark{a}} &
 \colhead{Dec\tablenotemark{a}} & \colhead{period} & \colhead{class.} \\
\colhead{W\#} & \colhead{} & \colhead{$''$} & \colhead{$''$} & \colhead{} &
 \colhead{\%} & \colhead{} & \colhead{} & 
\colhead{} & \colhead{} & \colhead{} & \colhead{(days)} 
}
\startdata

     12 & 1 &$-$0.18 &    0.11 & 1.40 & 0.401 & 18.96 &    1.07 &  0.98 & 00:24:09.2475 & $-$72:05:04.546 & 1.128   & AB \\
     14 & 1 &   0.04 &    0.08 & 0.76 & 0.070 & 17.61 &    0.87 &  0.89 & 00:24:08.4775 & $-$72:05:07.420 & 5.02    & AB \\
     15 & 1 &$-$0.01 &    0.03 & 0.42 & 0.405 & 22.47 & $-$0.94 &   \nd & 00:24:08.1886 & $-$72:05:00.166 & 0.17640 & CV \\
     18 & 1 &$-$0.11 & $-$0.10 & 1.05 & 0.205 & 21.17 &    2.10 &  1.45 & 00:24:07.8785 & $-$72:05:13.195 & 0.528   & AB \\
     26 & 1 &$-$0.00 &    0.08 & 0.70 & 0.058 & 22.29 &     \nd &  1.95 & 00:24:06.5664 & $-$72:05:12.285 & 0.39386 & AB \\
     27 & 1 &   0.02 & $-$0.00 & 0.64 & 0.018 & 23.11 & $-$1.26 &  2.26 & 00:24:06.0983 & $-$72:04:42.868 & 0.160   & CV \\
W\tablenotemark{d} & 1 &$-$0.03 &    0.06 & 0.83 & 0.326 & 22.30 &    1.50 &  0.70 & 00:24:05.7827 & $-$72:04:48.953 & 0.13294 & MSP \\
     30 & 1 &   0.00 & $-$0.07 & 1.82 & 0.220 & 20.44 & $-$0.71 &  1.16 & 00:24:05.7208 & $-$72:04:56.015 &\nd\tablenotemark{c}&CV \\
     34 & 1 &$-$0.13 &    0.13 & 1.98 & 2.432 & 22.75 & $-$0.12 &  1.03 & 00:24:04.9369 & $-$72:04:46.589 & 0.06767 & CV/MSP \\
     36 & 1 &   0.03 & $-$0.11 & 1.08 & 0.879 & 17.40 &    0.33 &  0.96 & 00:24:04.6452 & $-$72:04:55.300 & 1.108   & CV \\
     38 & 1 &$-$0.06 & $-$0.05 & 0.89 & 0.045 & 17.25 &    0.71 &  0.86 & 00:24:04.6382 & $-$72:04:46.007 & 1.866   & AB \\
     41 & 1 &   0.02 & $-$0.04 & 0.43 & 0.020 & 20.20 &    1.60 &  1.23 & 00:24:04.0533 & $-$72:05:01.221 & 0.4145  & AB \\
     42 & 1 &$-$0.00 &    0.01 & 0.37 & 0.004 & 19.88 & $-$0.67 &  0.36 & 00:24:03.9773 & $-$72:04:57.885 &\nd\tablenotemark{c}&CV \\
     44 & 1 &   0.01 & $-$0.11 & 1.34 & 0.806 & 23.20 & $-$1.05 &  2.46 & 00:24:03.4124 & $-$72:04:58.856 &     \nd & CV \\
     47 & 1 &$-$0.02 & $-$0.01 & 0.37 & 0.003 & 18.39 &    0.74 &  0.81 & 00:24:03.1840 & $-$72:05:05.123 & 0.5305  & AB \\
     72 & 1 &$-$0.01 & $-$0.07 & 0.53 & 0.043 & 17.40 &    1.10 &  0.99 & 00:24:10.1782 & $-$72:04:59.780 & 6.71    & AB \\
     73 & 1 &$-$0.15 &    0.20 & 1.89 & 0.559 & 17.33 &    0.59 &  0.79 & 00:24:06.9408 & $-$72:04:57.384 & 2.43    & AB \\
     75 & 1 &   0.35 & $-$0.03 & 2.30 & 1.115 & 16.98 &    0.62 &  0.79 & 00:24:06.0691 & $-$72:04:52.698 & 4.36    & AB \\
      1 & 2 &$-$0.01 &    0.03 & 0.32 & 0.019 & 20.37 & $-$1.07 &  0.36 & 00:24:16.7173 & $-$72:04:27.124 & 0.2405  & CV \\
      2 & 2 &$-$0.02 & $-$0.03 & 0.45 & 0.027 & 21.50 &    0.04 &  1.91 & 00:24:15.6337 & $-$72:04:36.312 &0.262\tablenotemark{b}&CV\\
      3 & 2 &$-$0.10 &    0.11 & 1.14 & 0.035 & 17.51 &    1.25 &  1.16 & 00:24:14.9180 & $-$72:04:43.454 & 5.34    & AB \\
      8 & 2 &   0.00 & $-$0.14 & 1.31 & 0.288 & 21.72 & $-$0.67 &  1.61 & 00:24:10.5048 & $-$72:04:25.579 & 0.11927 & CV \\
U\tablenotemark{d}&2&$-$0.04&0.02&0.26& 0.048 & 20.80 & $-$0.03 &$-$0.12& 00:24:09.5995 & $-$72:03:59.607 & 0.432 & MSP\\
     23 & 2 &   0.10 &    0.05 & 1.29 & 0.020 & 18.08 &    0.46 &  0.66 & 00:24:07.5283 & $-$72:04:41.533 & 0.25744 & AB \\
     33 & 2 &   0.14 &    0.19 & 1.45 & 0.825 & 21.40 & $-$0.68 &   \nd & 00:24:05.1432 & $-$72:04:21.531 &     \nd & CV \\
     45 & 2 &$-$0.02 & $-$0.01 & 0.27 & 0.012 & 21.91 & $-$0.98 &   \nd & 00:24:03.5197 & $-$72:04:22.856 &     \nd & CV \\
     69 & 2 &   0.11 &    0.16 & 0.84 & 0.063 & 18.31 &    0.81 &  0.96 & 00:24:12.3851 & $-$72:04:22.230 & 3.06    & AB \\
     70 & 2 &   0.16 &    0.27 & 1.86 & 1.438 & 23.30 & $-$0.42 &   \nd & 00:24:11.6261 & $-$72:04:44.077 &     \nd & CV \\
     76 & 2 &$-$0.31 &    0.10 & 1.43 & 0.177 & 17.34 &    0.52 &  0.78 & 00:24:06.2042 & $-$72:04:30.219 & 0.633   & AB \\
     92 & 2 &$-$0.04 & $-$0.06 & 0.29 & 0.008 & 16.16 &    0.45 &  0.28 & 00:24:15.4300 & $-$72:04:41.230 & 1.34    & AB \\
    167 & 2 &   0.03 & $-$0.20 & 0.84 & 0.070 & 16.62 &    0.61 &  0.42 & 00:24:09.9704 & $-$72:04:01.465 & 0.37212 & AB \\
     66 & 3 &   0.06 & $-$0.01 & 0.34 & 0.006 & 20.89 &     \nd &  1.34 & 00:24:15.7319 & $-$72:04:51.695 & 0.2108  & AB  \\
    163 & 3 &$-$0.08 &    0.01 & 0.38 & 0.009 & 17.02 &    0.40 &  0.61 & 00:24:24.9261 & $-$72:05:07.236 & 0.3450  & AB  \\
    182 & 3 &$-$0.19 &    0.49 & 2.77 & 0.369 & 19.99 &    1.54 &  1.14 & 00:24:19.1672 & $-$72:04:31.339 & 0.4174  & AB  \\
      9 & 4 &$-$0.09 &    0.08 & 0.64 & 0.055 & 17.04 &    0.65 &  0.74 & 00:24:10.2673 & $-$72:05:06.501 & 6.13    & AB  \\
     21 & 4 &$-$0.03 & $-$0.00 & 0.22 & 0.032 & 20.89 &    0.15 &  0.27 & 00:24:07.4969 & $-$72:05:27.196 & 0.07223 & CV \\
     22 & 4 &$-$0.11 & $-$0.15 & 1.09 & 0.129 & 18.08 &    0.61 &  0.82 & 00:24:07.5631 & $-$72:05:24.292 & 2.4494  & AB  \\
     25 & 4 &   0.02 & $-$0.04 & 0.29 & 0.014 & 21.32 & $-$0.68 &  1.23 & 00:24:06.8635 & $-$72:05:45.663 &\nd\tablenotemark{c}&CV \\
     59 & 4 &   0.02 &    0.01 & 0.11 & 0.002 & 20.08 &     \nd &  1.08 & 00:24:00.3385 & $-$72:05:54.156 & 0.24302 & AB  \\
     68 & 4 &$-$0.32 & $-$0.09 & 1.35 & 0.412 & 18.16 &    0.49 &  0.71 & 00:24:13.2434 & $-$72:05:23.372 & 1.118   & AB  \\
     94 & 4 &   0.05 & $-$0.12 & 0.61 & 0.065 & 21.67 &     \nd &  1.47 & 00:24:11.5193 & $-$72:05:14.533 & 0.27259 & AB  \\
    120 & 4 &   0.19 & $-$0.01 & 0.94 & 0.275 & 22.74 & $-$0.30 &  1.33 & 00:24:10.8120 & $-$72:06:19.926 & 0.220?  & CV \\
    121 & 4 &$-$0.08 &    0.11 & 0.62 & 0.070 & 21.78 &     \nd &  1.44 & 00:24:04.9081 & $-$72:06:04.256 & 0.15784 & AB  \\
    137 & 4 &$-$0.24 &    0.19 & 1.02 & 0.354 & 19.23 &    0.83 &  0.91 & 00:24:18.4695 & $-$72:06:02.872 & 0.5875  & AB  \\
    184 & 4 &$-$0.27 &    0.32 & 1.60 & 0.677 & 19.70 &    1.36 &  1.16 & 00:24:10.3112 & $-$72:05:29.076 & 0.7722  & AB  \\
\cutinhead{Marginal candidates}	  
        4? & 2 &   0.07 & $-$0.10 & 1.13 &   \nd &  17.50 &   0.61 &  0.83 & 00:24:13.2008 & $-$72:04:51.114 &     \nd & AB?  \\
T?\tablenotemark{d}&2&$-$0.10&$-$0.05&0.88&   \nd &$>$22.83&$<$0.95 &   \nd & 00:24:08.3049 & $-$72:04:38.786 &    & MSP? \\
       71? & 4 &$-$0.14 &    0.82 & 3.37 & 11.034&  20.88 &   0.37 &  0.26 & 00:24:10.3766 & $-$72:05:17.225 & 0.09877 & CV? \\
       93? & 4 &   0.53 & $-$0.28 & 2.61 & \nd   &  18.89 &   0.70 &  0.65 & 00:24:11.6536 & $-$72:05:07.832 & 0.2266  & AB?  \\
      140? & 4 &   0.43 & $-$0.17 & 2.09 & 4.125 &  23.16 & $-$0.65&  1.02 & 00:24:08.8636 & $-$72:05:43.491 &     \nd & CV? \\

\tablenotetext{a}{Coordinates were calculated using the STSDAS task METRIC
and archival image u5jm070cr.}
\tablenotetext{b}{X-ray period.}
\tablenotetext{c}{Significant non-periodic variability seen in $V$ and $I$ time series.}
\tablenotetext{d}{MSP from Camilo et al. (2000).}

\enddata
\label{tab.8267}
\end{deluxetable}

% ---- Table 4-------------------------------------------------------------------

%\tabletypesize{\scriptsize}
\tabletypesize{\tiny}

\begin{deluxetable}{rccrrcccrrccc}
\tablecolumns{13}
\tablewidth{0pc}
\tablecaption{\hst\ (GO-7503) data for counterparts to 47 Tuc \cha\ sources}
\tablehead{\colhead{Source} & \colhead{GO-8267} & \colhead{Chip \#} &
 \colhead{$\Delta$RA} & \colhead{$\Delta$Dec} & \colhead{signif.} & \colhead{chance} &
 \colhead{$F300W$} & \colhead{$F555W$} & \colhead{RA\tablenotemark{a}} &
 \colhead{Dec\tablenotemark{a}} & \colhead{class.} & \colhead{variability} \\
\colhead{W\#} & \colhead{} & \colhead{} & \colhead{$''$} & \colhead{$''$} & \colhead{} & \colhead{\%} &
  \colhead{} & \colhead{} & \colhead{} & \colhead{} & \colhead{} & \colhead{btw. epochs?} } 

\startdata

 15 & Y & 1 &$-$0.06 &    0.06 & 0.99 &  0.405 & 21.17 &    21.70 & 00:24:08.4485 & $-$72:05:00.495 & CV     & \\
 23 & Y & 1 &   0.06 &    0.10 & 1.56 &  0.020 & 18.57 &    18.12 & 00:24:07.7475 & $-$72:04:41.833 & AB     & \\
 27 & Y & 1 &   0.02 &    0.00 & 0.44 &  0.018 & 20.64 & $>$21.64 & 00:24:06.3484 & $-$72:04:43.184 & CV     & yes \\
 30 & Y & 1 &$-$0.01 & $-$0.06 & 1.29 &  0.220 & 19.90 &    20.92 & 00:24:05.9737 & $-$72:04:56.326 & CV     & yes \\
 34 & Y & 1 &$-$0.14 &    0.13 & 1.95 &  2.432 & 23.14 & $>$21.75 & 00:24:05.1874 & $-$72:04:46.899 & MSP/CV & \\
 36 & Y & 1 &   0.03 & $-$0.11 & 1.07 &  0.879 & 17.68 &    17.50 & 00:24:04.8959 & $-$72:04:55.599 & CV     & yes \\
 41 & Y & 1 &   0.01 & $-$0.05 & 0.43 &  0.020 & 22.03 &    20.01 & 00:24:04.3050 & $-$72:05:01.513 & AB     & \\
 42 & Y & 1 &$-$0.00 &    0.01 & 0.25 &  0.004 & 18.10 &    19.81 & 00:24:04.2273 & $-$72:04:58.184 & CV     & \\
 43 &   & 1 &   0.09 & $-$0.35 & 2.83 &  \nd   & 18.49 &    17.10 & 00:24:04.1995 & $-$72:04:43.607 & AB     & \\
 44 & Y & 1 &   0.00 & $-$0.11 & 1.34 &  0.806 & 20.82 & $>$21.56 & 00:24:03.6633 & $-$72:04:59.147 & CV     & yes \\
 47 & Y & 1 &$-$0.02 & $-$0.01 & 0.41 &  0.003 & 19.04 &    18.31 & 00:24:03.4351 & $-$72:05:05.413 & AB     & \\
 49 &   & 1 &$-$0.11 & $-$0.07 & 1.13 &  1.044 & 23.40 & $>$22.02 & 00:24:03.0707 & $-$72:04:47.501 & CV     & \\
 51 &   & 1 &   0.04 & $-$0.02 & 0.48 &  0.130 & 22.12 & $>$21.69 & 00:24:02.7856 & $-$72:04:49.258 & CV     & yes \\
 73 & Y & 1 &$-$0.17 &    0.22 & 2.04 &  0.559 & 17.92 &    17.40 & 00:24:07.1965 & $-$72:04:57.699 & AB     & \\
 75 & Y & 1 &   0.34 & $-$0.02 & 2.20 &  1.115 & 17.44 &    17.01 & 00:24:06.3214 & $-$72:04:53.012 & AB     & \\
 76 & Y & 1 &$-$0.29 &    0.14 & 1.46 &  0.177 & 17.93 &    17.38 & 00:24:06.4159 & $-$72:04:30.517 & AB     & \\
 82 &   & 1 &   0.10 &    0.22 & 1.32 &  2.286 & 21.50 & $>$21.47 & 00:24:01.3734 & $-$72:04:41.964 & CV     & \\
  9 & Y & 2 &$-$0.10 &    0.12 & 1.04 &  0.055 & 17.59 &    17.07 & 00:24:10.4786 & $-$72:05:06.823 & AB     & \\
 14 & Y & 2 &   0.08 &    0.06 & 0.92 &  0.070 & 18.55 &    17.54 & 00:24:08.7139 & $-$72:05:07.850 & AB     & \\
 21 & Y & 2 &$-$0.04 &    0.02 & 0.56 &  0.032 & 20.61 &    21.25 & 00:24:07.7102 & $-$72:05:27.552 & CV     & \\
 22 & Y & 2 &$-$0.12 & $-$0.13 & 1.51 &  0.129 & 18.52 &    18.04 & 00:24:07.7772 & $-$72:05:24.632 & AB     & \\
 25 & Y & 2 &$-$0.02 & $-$0.04 & 0.88 &  0.014 & 20.04 &    21.43 & 00:24:07.0828 & $-$72:05:46.041 & CV     & \\
 68 & Y & 2 &$-$0.35 & $-$0.07 & 1.90 &  0.412 & 18.75 &    18.18 & 00:24:13.4607 & $-$72:05:23.717 & AB     & \\
137 & Y & 2 &$-$0.34 &    0.20 & 1.92 &  0.354 & 20.24 &    19.17 & 00:24:18.6980 & $-$72:06:03.308 & AB     & \\
184 & Y & 2 &$-$0.29 &    0.33 & 1.97 &  0.677 & 21.60 &    19.83 & 00:24:10.5264 & $-$72:05:29.422 & AB     & \\
 56 &   & 3 &$-$0.03 &    0.02 & 1.53 &  0.026 & 22.58 &    23.32 & 00:24:02.0617 & $-$72:05:42.315 & CV     & yes \\
122 &   & 3 &   0.11 & $-$0.09 & 1.52 &  0.172 & 19.02 &    20.71 & 00:24:03.7760 & $-$72:06:21.924 & CV     & yes \\
 53 &   & 4 &   0.02 & $-$0.02 & 0.32 &  0.024 & 22.00 & $>$21.68 & 00:24:02.4777 & $-$72:05:11.466 & CV     & yes \\
 58 &   & 4 &$-$0.01 &    0.00 & 1.10 &  0.004 & 23.88 &    21.90 & 00:24:00.9036 & $-$72:04:53.415 & qLMXB  & yes \\
 64 &   & 4 &   0.02 & $-$0.05 & 1.34 &  \nd   & 22.47 &    21.48 & 00:23:57.6207 & $-$72:05:02.146 & AB     & \\
 85 &   & 4 &   0.03 &    0.11 & 0.73 &  0.203 & 21.29 &    19.48 & 00:23:59.3432 & $-$72:04:38.738 & CV     & \\				             	   			       				 
\cutinhead{Marginal candidates}	             	   			       				 
 31? & Y & 1 &   0.15 &    0.12 & 1.68 &  2.114 & 22.84 & $>$21.64 & 00:24:05.6108 & $-$72:05:04.609 & MSP/CV?    & \\
 35? & Y & 1 &   0.00 & $-$0.03 & 0.23 &  0.050 & 21.14 &    20.43 & 00:24:04.9965 & $-$72:05:06.456 & CV?    & \\
 55? &   & 1 &   0.20 & $-$0.41 & 2.39 & 10.824 & 21.65 &    22.80 & 00:24:02.1545 & $-$72:04:50.426 & CV?    & \\
 71? & Y & 2 &$-$0.17 &    0.84 & 4.06 & 11.034 & 20.93 &    21.37 & 00:24:10.5916 & $-$72:05:17.553 & CV?    & \\
140? & Y & 2 &   0.40 & $-$0.13 & 2.66 &  4.125 & 21.57 &    23.33 & 00:24:09.0801 & $-$72:05:43.873 & CV?    & \\

\tablenotetext{a}{Coordinates were calculated using the STSDAS task METRIC
and archival image u5470101r.}
				
\tablecomments{(1) 9 sources were in both the GO-7503 and 
GO-8267 FoVs, and have plausible IDs listed in Table \ref{tab.8267}, but
were too crowded in the GO-7503 images to be included here. These sources
are: W12, W18, W26, W29, W38, W59, W66, W94 and W121.\\  
(2) W120 was found very near the edge of the WF2 chip and its photometry is
not included here.\\
(3) 22 sources were found in the GO-7503 FoV, but were outside the GO-8267
FoV and had no plausible optical counterparts. There are 6 PC1 sources
(W40, W54, W57, W77, W79, W80), one WF2 source (W169), 7 WF3 sources (W52,
W78, W84, W90, W101, W102, W145), and 8 WF4 sources (W60, W61, W62, W65,
W81, W86, W87, W103). W77 is the X-ray counterpart to 47~Tuc~F and 47~Tuc~S. }
				
\enddata			
\label{tab.7503}
\end{deluxetable}		

% ---- Table 5-------------------------------------------------------------------

\tabletypesize{\scriptsize}

\begin{deluxetable}{rcrrccl}
\tablecolumns{7}
\tablewidth{0pc}
\tablecaption{\hst\ (GO-8267) data for blue variables}
\tablehead{\colhead{Name} & \colhead{$V$} & \colhead{$U-V$} & \colhead{$V-I$} 
 & \colhead{RA} & \colhead{Dec} & \colhead{period\tablenotemark{a}} \\
 \colhead{} & \colhead{} & \colhead{} & \colhead{} &
 \colhead{} & \colhead{} & \colhead{(days)} 
}
\startdata

1V36 & 17.69 &    0.30 &    0.11 & 00:24:06.2184 & $-$72:04:52.168 & 0.7944 \\
2V08 & 21.43 & $-$1.44 & $-$0.42 & 00:24:15.1166 & $-$72:03:36.569 & 0.4011 \\
2V30 & 21.36 & $-$1.20 &    0.38 & 00:24:14.8416 & $-$72:03:44.793 & 2.35 \\
3V06 & 22.90 & \nodata &    0.71 & 00:24:25.4577 & $-$72:04:38.204 & 0.2316 \\
3V07 & 19.63 & $-$0.01 & $-$0.06 & 00:24:14.2860 & $-$72:04:54.745 & 0.3163 \\
4V05 & 18.67 &    0.19 &    0.17 & 00:24:14.6191 & $-$72:05:40.371 & 0.2588 \\
 71? & 20.88 &    0.37 &    0.26 & 00:24:10.3766 & $-$72:05:17.225 & 0.09877 \\

\tablenotetext{a}{The orbital period assuming that the observed variations 
are ellipsoidal}

\enddata
\label{tab.bluevar}
\end{deluxetable}

\end{document}